\documentclass[numberedappendix]{emulateapj}

\newcommand{\kd}{$(u-B)_{\rm K}$}

\slugcomment{Accepted in AJ}

\begin{document}

\title{The BHB stars in the Survey Fields of Rodgers et al. (1993):
        New Observations and Comparisons with other Recent Surveys.  }

\author{T.D.Kinman\altaffilmark{1}          }
\affil{NOAO, P.O.Box 26732, Tucson, Arizona 85726, USA}

\author{ Warren  R. Brown}                             
\affil{Smithsonian Astrophysical Observatory, 60 Garden St.,    
 Cambridge, MA 02138, USA}

\altaffiltext{1}{ 
 The NOAO are operated by AURA, Inc.\ under cooperative 
agreement with the National Science
Foundation.}

\begin{abstract}
 We study blue horizontal branch (BHB) and RR Lyrae stars in the Rogers et  
 al.~(1993a) fields and compare their velocity and density distributions with 
 other surveys in the same part of the sky. 
 Photometric data are given for 176 early-type stars in the northern field.
 We identify fourteen BHB stars and four possible BHB stars, and determine
 the selection efficiency of the $Century$ Survey, the $HK$ Survey, and the 
 $SDSS$ survey for BHB stars. We give light curves and $\gamma$ radial
 velocities for three type $ab$ RR Lyrae stars in the northern field;
 comparison with the nearby $LONEOS$ Survey shows that there is likely to be 
 an equal number of lower-amplitude type $ab$ RR Lyrae stars that we do not 
 find. There are therefore at least {\it twice} as many BHB stars as type
 $ab$ RR Lyrae stars in the northern field --- similar to the ratio in the
 solar neighborhood.
 The velocity distribution of the southern field shows no evidence for an 
 anomalous thick disk that was found by Gilmore et al. (2002); the halo 
 velocity peaks at a slightly prograde rotational velocity but there is also 
 a significant retrograde halo component in this field. The velocity 
 distribution in the northern field shows no evidence of Galactic rotation
 for $|Z|$$\geq$ 4 kpc and a slight prograde motion for 
  $|Z|$$<$ 4 kpc. 
 The space densities of BHB stars in the northern field agree with an 
 extrapolation
 of the power-law distribution recently derived by de Propris et al. (2010). 
  For $|Z|$ $<$
 4 kpc, however, we observe an excess of BHB stars compared with this 
 power-law.  We conclude that these BHB stars mostly belong to a spatially 
 flattened, non-rotating inner halo component of the Milky Way in
 confirmation of the Kinman et al.~(2009) analysis of $Century$ Survey BHB 
 stars.

\end{abstract}


\section{Introduction}

  In this paper we identify blue horizontal branch (BHB) and RR Lyrae stars
  among the 176 early type stars in the Rodgers et al.\ (1993a) fields and
  compare their velocity and density distributions with those of other
  surveys in the same part of the sky. BHB and RR Lyrae stars are interesting
  because they are luminous tracers of the old stellar population. Our targets
  come from Rodgers et al. (1993a), who used the Anglo-Australian Observatory
  Schmidt Telescope to make an objective-prism survey (600 \AA\ mm$^{-1}$ at
  H$\gamma$) in two 70 deg$^{2}$ fields: $l = 90\arcdeg, b = -45\arcdeg$ in the 
  North (hereafter $NR$) and $l = 270\arcdeg, b =-45\arcdeg$ in the South
  (hereafter $SR$). They determined spectral types using the relative strengths
  of the Ca{\sc ii} K-line and the H$\delta$ Balmer line for stars in the 
  magnitude range $10 < V < 15.5$. They obtained follow-up coud\'{e} spectra
  (2.4 \AA~resolution) of 332 of these stars at the Mt.\ Stromlo 74-inch 
  telescope. The resulting sample of high Galactic latitude {\it metal-rich} 
  stars are discussed in Rodgers et al.\ (1993b), 
   an extension of Rodgers (1971).
  The sample of {\it metal-weak} stars is discussed in Rodgers et al.\ (1993c); 
  they conclude that this sample contains BHB stars but do not identify these 
  stars individually.

  We use photometry and spectroscopy to determine the stellar nature of the BHB and
  RR Lyrae candidates given by Rodgers et al.\ (1993c).  BHB stars can be identified
  with considerable certainty, for example, by using a $(B-V)$ {\it vs} $(u-B)_{K}$
  plot (Kinman et al.\ 1994). {\it GALEX} ultraviolet magnitudes can also be used to
  identify BHB stars (Kinman et al.\ 2007b). Rodgers et al. (1993a) give no colors for 
  their stars, only $V$ magnitudes calculated from the {\it Hubble Space Telescope} ($HST$)
  Guide Star Catalog (Lasker et al.\ 1990). We therefore obtain $V$, $(B-V)$ and
  $(u-B)_K$ photometry for most of the stars of spectral type A8 and earlier in the
  $NR$ field. Photometry is also taken from the 2MASS and {\it GALEX} catalogs. 
  In addition,
  the $NR$ field is covered by the Northern Sky Variability Survey (hereafter $NSVS$;
  W\'{o}zniak et al.\ 2004) and the {\it ASAS-3} variability survey 
   (Pojma\'{n}ski, 2002). We
  use these two surveys to study variability among the early-type stars and identify 
  possible new RR Lyrae stars. We obtain $V$ light-curves and phase-corrected radial
  velocities for the three brightest RR Lyrae stars in the $NR$ field. 

  The radial velocity distribution of the stars is important because the $NR$ and $SR$ 
  fields have Galactic longitudes for which the sight-line component of Galactic
  rotation is the greatest and thus for which radial velocity can distinguish between
  disk and halo populations. An additional constraint on the nature and origin of the
  BHB and RR Lyrae stars is provided by their spatial distribution. We combine our
  photometry and luminosity estimates to make accurate distance estimates and establish
  that most of these stars belong to a flattened, non-rotating inner halo 
  component of the Milky Way.

  \subsection{Comparison with Other Surveys}


 Gilmore et al.\ (2002) obtained radial velocities for $\sim$2000 main sequence {\it  
 turn-off} stars in two southern fields, one of which is coincident with (but smaller) 
 than the $SR$ field.
 The distribution of radial velocities in the Gilmore et al.\ fields showed
 a thick disk that had an unusually low galactic rotation at a few kpc above
 the plane; their data was also ``suggestive of a retrograde halo stream".
 They concluded that their sample was dominated by stars from a disrupted
 satellite that had merged with the disk of the Milky Way some 10 to 12 Gyr ago. 
 Navarro et al. (2004) suggested that these kinematically anomalous thick disk stars
 are connected with the
 metal-poor high-velocity ``Arcturus Group" in the solar neighborhood.
Carollo et al.\ (2010) have also identified a disk population with an 
anomalously low rotation (the metal-weak thick disk) and noted its similarity
to the population found by Gilmore et al.\footnote{The analysis of Carollo et
al. (2010) has, however, been criticized by Sch\"{o}nrich et al.\ (2010).}
 A broader discussion of these streams
has been given by Minchev et al. (2009). 

 Our main interest in this paper is the $NR$ field, which is overlapped by and/or 
 contiguous with a number of recent surveys.  
 Stripes 76 and 79 of the Seventh  Data Release ($DR7$) of the Sloan Digital Sky Survey $(SDSS)$
 (Abazajian et al.\ 2009) cross the $NR$ field and contain 33 stars 
  whose positions coincide with those
 listed in the $NR$ field, but only 8 of these (roughly those with $V$ $\geq$
 14.5) are faint enough to have reliable $SDSS$ photometry\footnote{
  The BHB star surveys by Xue et al. (2008) and Brown et al.\ (2010) that are
  based on the $DR6$ of the $SDSS$  do 
  not include stars from the $NR$ field.}.  Smith et al.\ (2010) have given 
 probabilities that a star is a BHB star for 27,074 stars in the $DR7$ of the
  $SDSS$; five  of our candidate BHB stars in the $NR$ field are included 
 in their catalog.
 The $NR$ field is also contiguous to the $SDSS$ Stripe 82 which has been searched
 for RR Lyrae stars, subdwarfs and BHB stars by Watkins et al.\ (2009), Smith et
 al.\ (2009) and Deason et al.\ (2010) respectively.
  These studies and that of Belokurov et al.\ (2007) 
 show that the $NR$ field is unlikely to be affected by the Hercules-Aquila
 Cloud (the nearest large sub-structure in the halo). 

 The two surveys most relevant to our study are the $HK$ Survey and the $Century$ Survey.
 Beers et al.\ (2007b) (hereafter $B2M$)  used 2MASS colors to assign ``High", 
 ``Medium' or ``Low" probabilities 
 that the stars of the $HK$ Survey (Beers et al.\ 1988, 1996) are BHB stars.
  The $B2M$ survey overlaps $\sim$50\% of the area
 of the $NR$ survey. In this overlap region they list 45 stars; 28 of these  
 were given  by Rodgers et al.\ (1993a) 
 and which we include in our list of BHB candidates
 (Table 5). The 17 stars not listed by Rodgers et al.\ are given separately in 
 Table 10 in  Appendix B; these are mostly stars that are either too red or too 
 faint to have been included in the $NR$ survey.
  The Century Halo Star Survey (Brown et al.\ 2004, 2008) (hereafter $CHSS$) 
 overlaps $\sim$30\% of the $NR$ field and has 12 stars in this area\footnote{This 
survey used the 2MASS survey to extract stars with 12.5$<$$J$$<$15.5,
 --0.20$<$$(J-H)$$<$0.10 and $-0.10<$$(H-K)$$<$0.10. Spectroscopic analysis 
 showed that 47\% of these were BHB stars.}.
 Ten are listed by Rodgers et al.\ (1993a). The other two have $V$$>$15.0 and thus   
 are fainter than the limit at which the $NR$ survey is complete.
  Both the $B2M$ and $CHSS$ surveys use $2MASS$ colors 
 to select BHB  candidates but since the $CHSS$ additionally uses slit spectra 
 (2.3\AA~resolution, $\lambda\lambda$ 3450---5450 \AA), it is much more 
 effective in identifying these stars.

  Spectra are already available for 18 of the stars in the $NR$ field; they
 were obtained with the 
  RC spectrograph at the 4m Mayall telescope at Kitt Peak ($\sim$0.8~\AA~resolution, 
 $\lambda$$\lambda$ 3880--4580 \AA). We are grateful to Dr.\ Nick Suntzeff 
 (priv.\ communication 1996, 1998) for making available the radial velocities from these
 spectra. Additional spectra for this study were obtained with the FAST 
 spectrograph of the Whipple 1.5-m telescope (2.3~\AA~resolution,  
 $\lambda\lambda$~3600--5500 \AA).

\section{Photometry of Stars in the $NR$ Field.}
 
 We first consider the photometry of the stars in the $NR$ field. We give
 photometric data for 176 stars in the Northern field of Rodgers et al.\ (1993)
   in Table 1 (at end of paper). New photometry is given for 105 of these stars;       
   Johnson $B,V$ photometry is taken from the literature for 7 others.

Our new photometric observations were made with the KPNO 0.9-m telescope between 
 August 1996 and December 1997, inclusive. The detector was a 
  512$\times$512 Tektronix chip under control of the {\it CCDPHOT} program
  (Tody \& Davis, 1992). The measurements were made in a 128$\times$128 pixel
  area at the center of the chip. This 90$\times$90 arcsec area was small 
  enough to allow a rapid readout and yet large enough for acquisition to
  be easy.
  The filter slide was under computer control so that the observations could
  be made with an assigned integration time for each filter and this cycle
  could be repeated for a chosen  number of times.
  Bias and flat-field observations were made in the usual way and used by the
  program to compute instrumental magnitudes that were immediately available
  at the end of each integration cycle. Thus, the approximate 
  magnitude and color of the star were available immediately after
  each integration cycle.  The integrations through the various filters 
  were made in rapid succession and so  the  colors were relatively
  unaffected by slow trends in the transparency. 
 We concentrated on the stars with  spectral types 
  earlier than A8.  One or two 
 integration cycles with the $CCDPHOT$ program were sufficient
  to decide whether a star had the appropriate $(B-V)$ color 
  for a BHB star; if it had, the observations were continued. 

 The $V$, $B-V$ and $(u-B)_{K}$ observations were made as 
 described in Kinman et al.\ (1994). We used the $B,V$ standards of 
 Landolt (1992) --- primarily those in SA-114 and SA-115. For \kd , the 
 following secondary standards near the Rodgers field were used:
 HD~2857 (2.094); BD +02--0089 (1.938); BD +00--0145 (1.831); SA 114--750 
 (1.952);  SA 115--271 (1.918); SA 115--420 (1.924). 
  The adopted value of \kd~ is given in parentheses 
 following the name of each star. We measured the Str\"{o}mgren H$\beta$
 index for 30 of the brighter program stars ($V <$14) using well-observed 
 early-type stars taken from Stetson (1991) as standards.
 The E$(B-V)$ correction for galactic extinction given in col.\ 7 of Table 1 
 is from Schlegel et al.\ (1998).  We assumed 
   A$(V)$ = 3.1$\times$E$(B-V)$ and  A$(u-B)_{K}$ = 0.89$\times$E$(B-V)$  
 (Kinman et al.\ 1994).

\begin{figure}
\includegraphics[width = 3.5in]{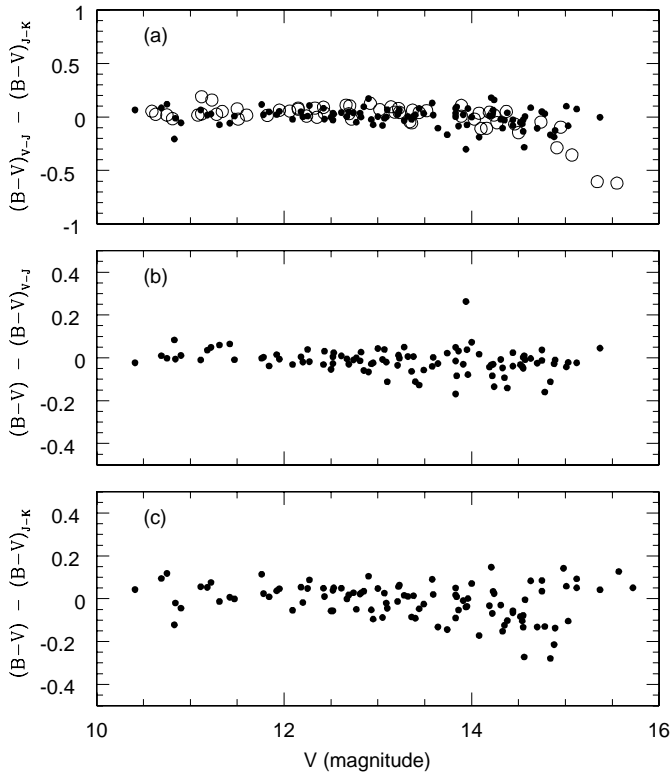}
\caption{
(a) The difference between the $(B-V)$ calculated from $(V-J)$ and that from
  $(J-K)$ $vs.$ $V$ magnitude. The filled circles are for the calibrating stars
  (with known $(B-V)$). The open circles are for stars whose $(B-V)$ was not 
  observed. 
(b) The difference between the observed $(B-V)$ and that calculated from 
    $(V-J)$ $vs.$ $V$ magnitude. 
(c) The difference between the observed $(B-V)$ and that calculated from 
    $(J-K)$ $vs.$ $V$ magnitude. 
\label{Fig2}}
\end{figure}

 Our mean $V$, $B-V$ and \kd~ for the program stars are given with their $rms$
 errors in cols.\ 3, 4 and 5 of Table 1. A $V$ magnitude that has no error 
 and a superscript~$^{a}$ is taken from the {\it ASAS-3} catalog
(Pojma\'{n}ski 2002). A $V$ with a superscript~$^{b}$ is a mean of
 the magnitudes 
 given in the {\it ASAS-3} catalog and the {\it GSC-2.3} catalog (Lasker et al.\ 2008).
 The sources of the other $V$ magnitudes are given in the table notes.
 
  The {\it GALEX} 
  $NUV$ magnitudes (col.\ 8) are taken from {\it MAST} (The Multimission Archive 
 at STSci, http://archive.stsci.edu/). $NUV$ is a near-$UV$  magnitude 
 (effective wavelength 2267\AA~) that can be used to select BHB stars 
 (Kinman et al.\ 2007b). The $J$ and $K$ magnitudes are taken from the 2MASS Point
 Source Catalog using the {\it Vizier} access tool. The $m_{R}$ red magnitude
 (col.\ 11) 
 and its scatter ($\Sigma_{mR}$) (col.\ 12) were taken from the Northern Sky Variability
 Survey (Wo\'{z}niak et al.\ 2004). 

  We also calculated   $(B-V)_{0}$ from both $(V-J)_{0}$ and from
 $(J-K)_{0}$ using the following relations that were derived from 
 the program stars with known  $(B-V)$: 
 \begin{equation}
    (B-V)_{0} = 0.588\pm0.022\times (V-J)_{0}  -0.041\pm0.011   \\ 
 \end{equation}
 \begin{equation}
    (B-V)_{0} = 1.553\pm0.084\times (J-K)_{0}  +0.031\pm0.013   \\ 
 \end{equation}
 Brown et al.\ (2008) derived $(B-V)_{0}$ from Balmer line equivalent widths but we
 found that the scatter in the relation between our $(B-V)_{0}$ and the Balmer line
 equivalent widths given in Rodgers et al.\ (1993b) is much larger than for the
 relations with $(V-J)_{0}$ and $(J-K)_{0}$. We therefore only used the relations
 with $(V-J)_{0}$ and $(J-K)_{0}$ (given above) to derive $(B-V)$. The scatter in 
 these relations is shown in Fig.\ \ref{Fig2}(b) and Fig.\ \ref{Fig2}(c) and the 
difference 
 between the
 $(B-V)$ derived from the two relations is shown in Fig.\ \ref{Fig2}(a). The $(V-J)$ 
relation
 was not used for stars with $V$ $>$ 14.5 because the $V$ magnitude is poorly 
 determined for fainter stars. For stars with $V$  $<$ 14.5, the color given in
 col.\ (3) of Table 1 is followed by a single colon and can be assumed to have an
 error of $\sim$0.1 magnitude. 
  For stars with $V$ $>$ 14.5, the colors are followed by
 a double colon and can be assumed to have an error of $\sim$0.2 mag.. Only a
  few of these stars (75 (Pn24l--13), 104 (Pn23l2--58), 132 (Pn24l--20) and  
 159 (Pn24l--60)) have colors that may be blue enough for them to be possible BHB 
 candidates. 

 The stars in Table 1 are given in order of
 RA and identified by a running number (col.\ 1) and the ID given by Rodgers et
 al.\ (1993a) (col.\ 2); the former is used throughout the rest of this paper. 
 The coordinates of the brighter stars given by Rodgers et al.\ (1993a) were 
 taken from the $HST$ 
 Guide Star Catalog and are accurate to one or two arcsec and are not repeated 
 here. The 
 coordinates of some of the fainter stars, however, were of lower accuracy and
 improved coordinates (taken from the USNO B Catalog) are given for them in 
 Table 9 in the Appendix B.1.

\begin{deluxetable*}{cccccccc}
\tablewidth{0in}
\tabletypesize{\footnotesize}
\setcounter{table}{2}
\tablecaption{ Stars showing evidence of Variability. } 
\tablehead{
\colhead{No.\tablenotemark{a}} &
\colhead{ID \tablenotemark{b}} &
\colhead{ RA     } &
\colhead{ DEC    } &
\colhead{ Spectral Type \tablenotemark{c} } &
\colhead{ Period } &
\colhead{ Type \tablenotemark{d}  } &
\colhead{ Notes  }\\ 
 & &\multicolumn{2}{c}{J~2000}&       &(days) &    &
}
\startdata
   12  &CS 29521-078  &23:10:43.0 &+10:36:03 & A0 &$\cdots$ &EC?      &  (1)   \\
   16  & IX PEG       &23:11:12.9 &+13:50:56 & A2 & 0.601 & RR{\it ab}&  (2)  \\
   37  & 14601441      &23:14:17.5 &+02:37:31 & A7 & 0.554 & EC        &  (3)   \\
   47  & 11843625      &23:15:46.0 &+09:18:01 & F0 & 0.381 & EC        &  (4)   \\
   84  & NSVS 11847482 &23:21:51.0 &+12:47:24 & A3 & 0.555 & RR{\it ab}&  (5)   \\
  164  & NSVS 14611789 &23:32:18.8 &+06:47:03 & F0 &$\cdots$ &$\cdots$ &  (6)   \\
  168  &CS 30333-0136  &23:32:59.0 &+08:26:14 & A3 &$\cdots$ &$\cdots$ &  (7)   \\
  169  & NSVS 9062655  &23:33:14.7 &+09:00:57 & A1 &$\cdots$ &$\cdots$ &  (8)  \\
  173  & NSVS 9063965  &23:34:58.9 &+13:44:06 & F0 &0.694  & RR{\it ab}&  (9)  \\ 
   A   & NSVS 9062108  &23:31:50.0 &+12:51:00 &$\cdots$&0.328& EC&(10)\\
\enddata

\tablenotetext{a}{Number given in Table 1 and Fig. 2. A is described in text.} 
\tablenotetext{b}{Identification } 
\tablenotetext{c}{Rodgers et al.\ (1993b) } 
\tablenotetext{d}{RR = RR Lyrae; EC = Contact binary.}
                                                           
\tablenotetext{(1)}{Norris et al.\ (1999) give $V$ = 13.65, $(B-V)$ 
= +0.34, $(U-B)$ = +0.01 
in their search for stars of low metal abundance but this star is not included in the
Beers et al.\ (2000) catalog of metal-weak stars. 
We found  $V$ = 13.58, $(B-V)$ = 0.334 and $(u-B)_{K}$ = +1.730; this suggests 
that the star is probably  a low-amplitude contact binary. }
\tablenotetext{(2)}{Wils et al.\ (2006) and Kinemuchi et al.\ (2006) 
give periods of
0.60099 and 0.60103 days respectively. Both periods are based on $NSVS$ data. 
Our photometric and spectroscopic observations are discussed in Sec. 4.}
\tablenotetext{(3)}{The period is from the catalog of Gettel et al.\ 
(2006) who list it
as a contact binary with a $V$  amplitude of 0.393 mag. }
\tablenotetext{(4)}{The period is from the catalog of Gettel et al.\ 
(2006) who list it
as a contact binary with a $V$  amplitude of 0.325 mag. }
\tablenotetext{(5)}{Wils et al.\ (2006) and Kinemuchi et al.\ (2006) 
give periods of
0.55515 and 0.55514 days respectively. Both periods are based on $NSVS$ data. 
Our photometric and spectroscopic observations are discussed in Sec. 4.}
\tablenotetext{(6)}{The evidence for variability from the $NSVS$ data 
is not supported by
that from the $ASAS-3$ catalog (Pojma\'{n}ski, 2002).}
\tablenotetext{(7)}{The evidence for variability from the $NSVS$ data 
is not supported by 
that from the {\it ASAS-3} catalog. 
 The catalog of HB and metal-weak
 stars (Beers et al.\ 2007a) gives $V$ = 12.18, $(B-V)$ = 0.23 and $(U-B)$ = +0.17
 compared with our $V$ = 12.20, $(B-V)$ = +0.234 and $(u-B)_{K}$ = 2.034. 
 We classify it as a BHB star.}
\tablenotetext{(8)}{The evidence for variability from the $NSVS$ data 
is not supported by 
that from the $ASAS-3$ catalog. We classify it as a BHB star.}
\tablenotetext{(9)}{Wils et al.\ (2006) and Kinemuchi et al.\ (2006) 
give periods of
0.69425 and 0.69417 days respectively. Both periods are based on $NSVS$ data. 
Our photometric and spectroscopic observations are discussed in Sec. 4.}
\tablenotetext{(10)}{The period is taken from the $ASAS-3$ catalog  
where the star 
is classified as RR{\it c}/EC. The 2MASS colors $(J-K)_{0}$ = 0.23 and 
$(V-J)_{0}$$\sim$1.0 are too red, however, for it to be an RR{\it c}. 
Spectra taken 
on 2009 Dec 19 UT with the FAST spectrograph of the Whipple Observatory 1.5-m
telescope (spectral resolution 2.3 \AA; $S/N$ = 30) show that this star and the
eclipsing star No. 37 had a FWHM for the $\lambda$4481 line of 2.4 and 2.6 \AA~
respectively. These FWHM are substantially larger than the FWHM of
1.1 and 1.2 \AA~ found for the RR Lyrae stars Nos. 16 and 18 respectively from
spectra taken on the same night. This line width difference strongly supports
the classification of this star as an eclipsing binary (Kinman \& Brown, 
2010).}

\end{deluxetable*}

\begin{figure}
\includegraphics[width = 3.5in]{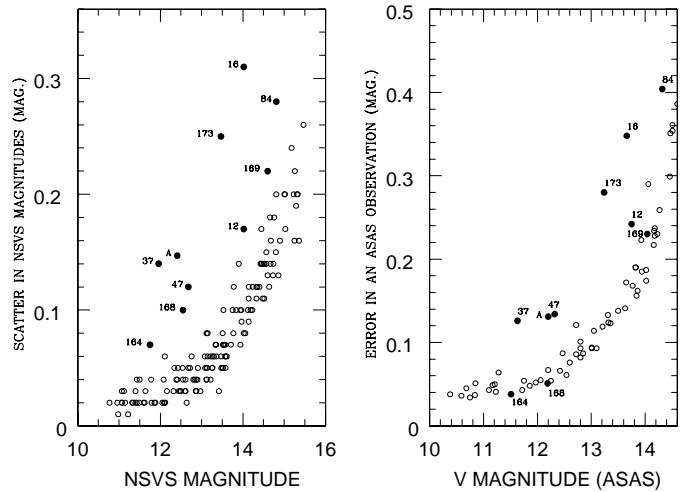}
\caption{(Left) The scatter ($\Sigma_{mR}$) in the individual $NSVS$ magnitudes
 ($m_{R}$) 
  (ordinate) {\it vs.} $m_{R}$. (Right) The error of a a single $V$ magnitude 
  in the $ASAS-3$ catalog {\it vs.} the $V$ magnitude. The significance of 
  the numbers is discussed in Sec.\ 3. 
\label{Fig3}}
\end{figure}

\section{ The Variability of the Program Stars in the $NR$ Field.}

 Before discussing the colors of our program stars, we need to test them
for  variability. Fig.\ 2 (left) shows a plot of the scatter in the
 individual magnitudes ($\Sigma_{mR}$) against their red magnitude ($m_{R}$) 
 as given in the Northern Sky Variability Survey (W\'{o}zniak et al.\ 2004).
 Most of these stars show a well-defined trend of increasing scatter with
 increasing $m_{R}$. Stars that  show more scatter than this trend (at roughly
 more than the 2$\sigma$-level) are shown by filled circles and labelled with the 
 number of the star (Table 1, col.\ 1). Fig.\ 2 (right) shows a similar plot of the
 error of a single observation in $V$ of a star in the $ASAS-3$ catalog
  (Pojma\'{n}ski,~2002) 
 against its $V$ magnitude. The numbered stars in Fig.\ 2 (left) are also shown by 
 numbered filled circles in Fig.\ 2 (right). The star marked A in both plots is NSVS
 9062108 (23:31:50, +12:51:00 (2000) ); it is listed in the $ASAS-3$ catalog as  type 
 RRC/EC with 12.05$<$$V$$<$12.43 and a period of 0.328712 days. More information 
 about these stars (including previous identifications as variables) is given in 
  the notes to Table 2. 
In summary, three  (16, 84 and 173) are RR{\it ab} stars,
four (12, 37, 47 and A) are eclipsing stars and the remaining three 
(164, 168 and 169) are probably not variable. We have made new photometric and 
spectroscopic observations of the three RR Lyrae stars and these are discussed in
Sec.\ 4.

The RR Lyrae star IO PEG  (period = 0.567 days; 
 15.0$< m_{pg} <$ 16.5; Goranskij, 1986) is too faint to be listed in the 
$NSVS$ and has only 3 observations in the $ASAS-3$ catalog ($<V>$ = 15.3$\pm$0.3).
Although its magnitude is uncertain, we take it to be fainter than the limit 
of the $NR$ survey.

\begin{figure*}
\includegraphics[width=2.5in,angle=-90]{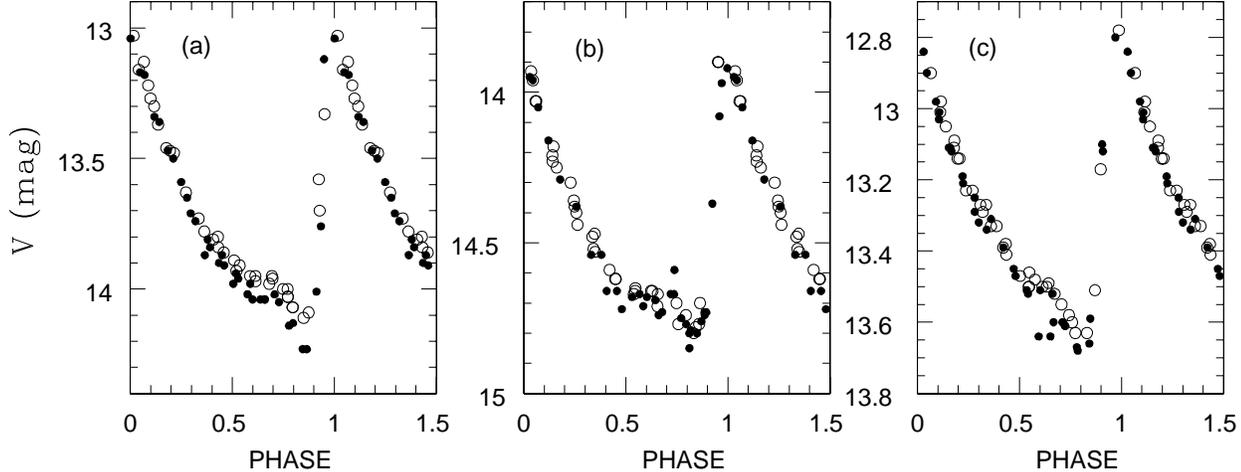}
\caption{ $V$-magnitude light curves for three type $ab$ RR Lyrae stars in the
 $NR$ field; (a) No. 16 (Pn23l2--1); (b) No. 84 (Pn23l2--38) and (c) No. 173
(Pn23l2--41). Filled circles are Tenagra observations in 2008 and open 
    circles are Tenagra observations in 2009.
\label{Fig 10}}
\end{figure*}

\begin{figure}
\includegraphics[width = 3.5in]{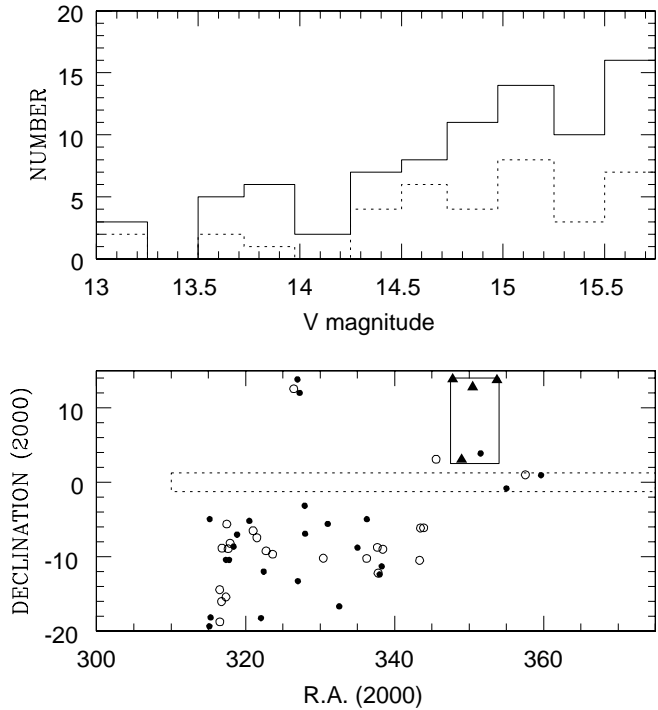}
\caption{(above) The number of RR Lyrae stars per 0.25 mag.\ interval in  400 
deg$^{2}$ of the $LONEOS$ survey near the $NR$ field. The full histogram shows
the total number and the dotted histogram those with a $V$-amplitude less than
 0.75 mag. (below) The location of the stars in the $LONEOS$ survey. 
  Those with $V$ $<$ 14.75 are shown by filled circles and those with 14.75 $\leq$
 $V$ $\leq$ 15.25 by open circles. The four $NSVS$ stars in the $NR$ field are 
 shown by filled triangles. The location of the $NR$ field is shown by the solid
 rectangle and the $SDSS$ Stripe 82 field is shown by the dotted rectangle.
\label{Fig 11}}
\end{figure}

\section{RR Lyrae stars.}

\subsection{RR Lyrae stars in $NR$}

   In addition to BHB stars, RR Lyrae stars are important tracers of the
   halo. In our discussion of the 
    variability of the stars in the $NR$ field in Sec. 3, we found
  that Nos 16, 84 and 173 (Table 2 and Fig. 2) are type $ab$ RR Lyrae stars.
  We made photometric observations of these stars in 2008 and 2009 with 
  the commercial robotic f/7 0.8-m telescope of the Tenagra Observatory in 
  Arizona which has a 1024$\times$1024 SITe CCD. Details of similar photometry
  with this telescope are given by Kinman \& Brown (2010). Periods were  
  determined from this data using the periodogram program of Horne
  \& Baliunas (1986). The period of 0.601089 days found for No. 16 (IX Peg) 
  is satisfactorily close to the periods of 0.60103 days and 0.60099 days 
  that were found by Kinemuchi et al.\ (2006) and Wils et al.\ (2006) 
  respectively from the Northern Sky Variability Survey ($NSVS$) data 
  (W\'{o}zniak et al.\ 2004). The periods of 0.555135 and 0.6945125 days found
  for Nos 84 and 173 respectively are also close to the periods of
  0.55514 and 0.69417 days that Kinemuchi et al.\ (2006) found for them. 
  The light curves for these stars are given in Fig.\ 3. Metallicities [Fe/H]
  were derived from the amplitudes and periods (using eqn.\ 6 in Sandage 
  (2004)) and are given with other photometric data in Table 3.

 Radial velocities were measured from spectra taken 
 in December 2009 with the FAST 
 spectrograph of the Whipple 1.5-m telescope (resolution 2.3~\AA~ and waveband 
 $\lambda\lambda$~3600--5500 \AA). The $\gamma$-velocities were derived 
 following Liu (1992). The spectra of Nos 16 and 84 were were taken close
 to maximum light and so were not suitable for deriving [Fe/H] from the
  Ca\,{\sc ii}  K-line equivalent width. The spectrum of No. 173 was taken 
 at phase 0.451, however, so we were able to derive an [Fe/H] of --1.99 by
 the $\Delta$S-method (Preston 1959). This [Fe/H] is in satisfactory 
 agreement with the --2.15 derived from the amplitude and period. The 
 spectroscopic data for these stars are summarized in Table 3.
 
  Kinemuchi et al.\ (2006) found a fourth and fainter type $ab$ RR Lyrae star 
  ($NSVS$ 14602495; R.A. 23:15:56.66, DEC. +03:02:58 (2000)) within the $NR$
  field but not identified by Rodgers et al.\ (1993a). 
    It has a period of 0.59355 days and a $V$ amplitude of 1.56 mag, from which
 they derived an [Fe/H] of --2.60. Our $V$ amplitudes are less than theirs by a
 factor of 0.70$\pm$0.06, so the amplitude of this star is 1.09 mag. on our scale. 
 This impllies an [Fe/H] of --1.9 according to Sandage (2004). 
 Kinemuchi et al.\ also give an intensity-weighted 
 $V$-magnitude of 15.03 mag. Their mean $V$-magnitudes are calculated from 
 $NSVS$ unfiltered CCD magnitudes and (for the other three RR Lyrae stars)
 average 0.26 mag fainter than the intensity-weighted $V$ magnitudes that we
 derived from our photometry. We have therefore assumed an intensity-weighted 
 $V$-magnitude of 14.77 for this star and an absolute $V$ magnitude of +0.45
 (following Clementini et al.\ (2003)) to derive a distance of 6.67 kpc. 
 
\subsection{RR Lyrae stars in $LONEOS$ Survey}

 The $LONEOS$ survey for RR Lyrae stars by Miceli et al.\ (2008) covers the 
 sky in the neighborhood of the $NR$ field; they
  discovered 838 type $ab$ RR Lyrae stars in this field. 
  Fig. 4\ (below) 
 shows the location of the stars in this survey that have $V$ $<$ 15.25 and that 
 have 21$^{h}$ $<$ R.A.$<$ 01$^{h}$ and declinations between --20$^{\circ}$ and +15$^{\circ}$;
 the $V$ magnitude distribution of the brighter of the $LONEOS$ variables is 
 shown in Fig.\ 4 (above). About 45\% of this sample have $V$ amplitudes of less 
 than 0.75 mag. One of these lower amplitude variables lies in the $NR$ field but 
 is not identifiable as a variable in the $NSVS$ because it lies close to a
 brighter star. 

 The sample of $LONEOS$ stars shown in Fig.\ 4 cover roughly 400 deg$^{2}$ and
 contains 31 stars with $V$ $\leq$ 14.75 and 42 stars with $V$ $\leq$ 15.00.
 We might therefore expect 5 and 7 stars in these magnitude intervals, 
 respectively, in the 70 deg$^{2}$ of $NR$.

\begin{deluxetable}{cccc}
\tablewidth{3.5in}
\tabletypesize{\scriptsize}
\setcounter{table}{3}
\tablecaption{Data for three RR Lyrae stars in the $NR$ field.}
\tablehead{ 
\colhead{ID \tablenotemark{a}} &
\colhead{ Star 16} & 
\colhead{ Star 84} & 
\colhead{Star 173}
}

\startdata
 $GCVS$          &     IX PEG    &  $\cdots$     &  $\cdots$         \\
 $NSVS$          & 11840238      &  11847482     &   9063965         \\ 
 RA(J2000)       & 23:11:12.84   & 23:21:50.95   & 23:34:58.88       \\ 
 DEC(J2000)      & +13:50:56.2   & +12:47:24.5   &  +13:44:06.3      \\  
 Type            &  $ab$         &   $ab$        &    $ab$           \\  
 Period (days)   & 0.601089      &  0.555135     &   0.694125        \\  
 JD(max)\tablenotemark{b} &2454776.9770    &2451444.6900    &2454777.5497   \\
$<V>$\tablenotemark{c} &13.676          &14.443          &13.269         \\
 $V_{max}$       & 13.030        &  13.900       & 12.780                    \\
 $V_{min}$       & 14.158        &  14.801       &    13.655                 \\
 Rise Time       & 0.160         &  0.175        &   0.220                   \\
 $\phi_{31}$     & 2.05          &  1.82         &   2.42                    \\
 E$(B-V)$        & 0.070         &  0.059        &   0.071            \\
 $[Fe/H]$\tablenotemark{d} &--2.02          &--1.41          &--2.15  \\
 $[Fe/H]$  \tablenotemark{e} &--2.43          & ...            &--2.16        \\
 JD(hel) \tablenotemark{f} &2455184.5558    &2455184.6159    &2455184.6201   \\
   T     \tablenotemark{g} &300             &270             &270            \\
  Phase  \tablenotemark{h} &  0.067         & 0.966          & 0.451         \\
 Rad. Vel. \tablenotemark{i} &--359.8$\pm$5.1 &--113.8$\pm$4.6 &--098.6$\pm$5.9  \\
 Rad. Vel. \tablenotemark{j} & --325.9        & --87.5         & --104.4     \\
$\Delta$S \tablenotemark{k} & ...            & ...            & 10.0         \\
 $[Fe/H]$  \tablenotemark{l} &  ...           &  ...           & --1.99        \\
 M$_{v}$ \tablenotemark{m} & 0.428          &  0.558         & 0.434         \\
 Distance\tablenotemark{n} & 4.01           &  5.47          & 3.31          \\
\enddata

\tablenotetext{a}{No. of star in Table 1.}  
\tablenotetext{b}{Heliocentric Julian Date of maximim light.}  
\tablenotetext{c}{Intensity-weighted mean $V$ magnitude.}  
\tablenotetext{d}{Derived from eqn. (6) in Sandage (2004). }  
\tablenotetext{e}{Kinemuchi et al.\ (2006).                                       }  
\tablenotetext{f}{Heliocentric Julian Date of mid-exposure of spectrum.          }  
\tablenotetext{g}{Integration time for spectrum in seconds.                      }  
\tablenotetext{h}{Phase of spectrum.                                             }  
\tablenotetext{i}{Observed radial velocity in km s$^{-1}$ relative to LSR.       }  
\tablenotetext{j}{$\gamma$-velocity in km s$^{-1}$ relative to LSR.              }  
\tablenotetext{k}{Preston (1959) $\Delta$S derived from the calibration of 
    equivalent widths given in Kinman \& Carretta (1992). }
\tablenotetext{l}{Derived from $\Delta$S following Suntzeff et al.\ (1994).       }  
\tablenotetext{m}{Absolute $V$ magnitude from $[Fe/H]$ following Clementini et 
  al.\ (2003). }
\tablenotetext{n}{Distance in kpc. }  
\end{deluxetable}

\begin{figure}
\rightline{\includegraphics[width = 3.0in]{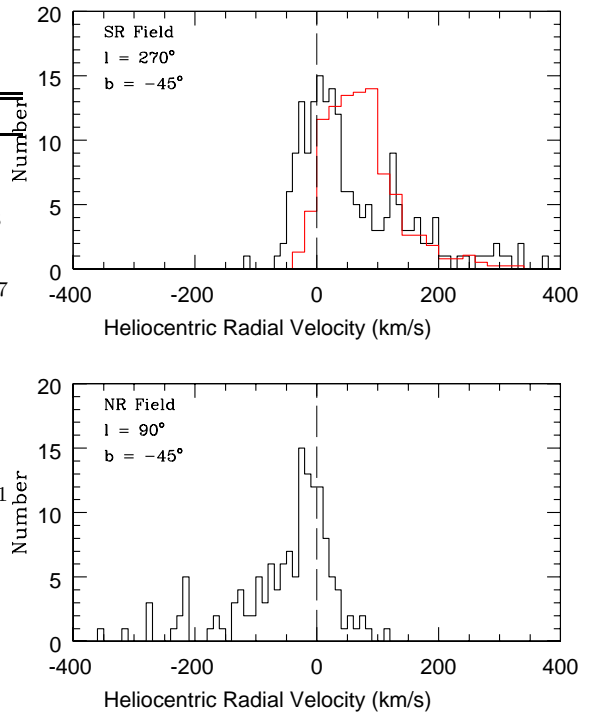}}
\caption{
  Distribution of the heliocentric radial velocities (V$_{hel}$) in km s$^{-1}$
  for the $SR$ (above) and $NR$ (below) fields are shown by the black 
  histograms for the early-type stars found by Rodgers et al.\ (1993).
 The red histogram shows the distribution of the turn-off stars 
 found in the (l,b) =
 (270\arcdeg, --45\arcdeg) field with $V$ $<$ 18 by 
  Gilmore et al.\ (2002). The total number of the turn-off stars 
  has been normalized 
 to equal that of the early-type stars.
\label{Fig1}}
\end{figure}

\begin{deluxetable*}{ccccccc}
\tablewidth{0cm}
\tabletypesize{\footnotesize}
\setcounter{table}{4}
\tablecaption{Comparison of $NR$ radial velocities with those from other sources.}
\tablehead{ 
\colhead{  No.  } &
\colhead{  ID \tablenotemark{a}  } &
\colhead{$NR_{LSR}$ \tablenotemark{b}} & 
\colhead{$NR_{helio}$ \tablenotemark{c}} & 
\colhead{RV$_{KPNO}$\tablenotemark{d}} &
\colhead{RV$_{Century}$\tablenotemark{e} } &
\colhead{RV$_{Adopted}$\tablenotemark{f} } \\
         &         & km s$^{-1}$  & km s$^{-1}$ & km s$^{-1}$ & km s$^{-1}$  &  km s$^{-1}$ \\
}
\startdata
        4 &    Pn24l--37   &   +015       &   +009    &   +019     &  $\cdots$   & +013  \\       
       16 &    Pn23l2--1   &  --352       &  --358    &  --351     &  $\cdots$   & --355  \\       
       38 &    Pn24l--51   &  --147       &  --153    &  --186     &  $\cdots$   & --171  \\       
       41 &    Pn23l2--14  &  --256       &  --262    &  --289     &  $\cdots$   & --277  \\       
       48 &    Pn24l--45   &  --130       &  --136    &  --104     &  $\cdots$   & --121  \\       
       60 &    Pn24l--42   &  --207       &  --213    &  --216     &  $\cdots$   & --216  \\       
       63 &    Pn23l2--2   &  --018       &  --024    & $\cdots$   &  --065      & --045  \\       
       72 &    Pn23l2--4   &   +014       &   +008    & $\cdots$   &  --002      & +002  \\       
       90 &    Pn23l2--36  &  --025       &  --031    & $\cdots$   &  --020      & --028  \\       
       99 &    Pn23l2--22  &  --201       &  --207    &  --232     &  --217      & --221  \\       
      110 &    Pn23l2--28  &  --061       &  --067    & $\cdots$   &  --0054     & --062  \\       
      115 &    Pn24l--55   &  --142       &  --148    &  --127     &  $\cdots$   & --140  \\       
      120 &    Pn24l--56   &  --047       &  --053    &  --086     &  $\cdots$   & --072  \\       
      126 &    Pn23l1--18  & $\cdots$     & $\cdots$  & $\cdots$   &  --057     & --059  \\       
      128 &    Pn23l2--30  &   +063       &   +057    &   +047     &  $\cdots$   & +050   \\       
      134 &    Pn24l--8    &  --117       &  --123    &  --140     &  $\cdots$   & --134  \\       
      140 &    Pn23l2--53  &  --038       &  --044    & $\cdots$   &  --051      & --050  \\       
      141 &    Pn24l--22   &  --116       &  --122    &  --123     &  $\cdots$   & --125  \\       
      145 &    Pn24l--4    &  --131       &  --137    &  --174     &  $\cdots$   & --158  \\       
      146 &    Pn24l--27   &  --038       &  --044    &  --066     &  $\cdots$   & --057  \\       
      148 &    Pn24l--26   &  --070       &  --076    & $\cdots$   &  --027      & --054  \\       
      151 &    Pn24l--6    &   +043       &   +037    &   +045     &  $\cdots$   & +039  \\       
      157 &    Pn23l1--39  &  --056       &  --062    &  --051     &  --0052     & --057  \\       
      158 &    Pn24l--28   &   +042       &   +036    &   +033     &  $\cdots$   & +032  \\       
      163 &    Pn23l2--43  &   +015       &   +009    & $\cdots$   &    000      & +002  \\       
      169 &    Pn23l2--45  &    000       &  --006    &  --030     &  $\cdots$   & --020  \\       
\enddata

\tablenotetext{a}{ID from Rodgers et al.\ (1993b)}  
\tablenotetext{b}{Radial velocity relative to LSR. Rodgers et al.\ (1993b)}  
\tablenotetext{c}{Heliocentric radial velocity. Rodgers et al.\ (1993b)} 
\tablenotetext{d}{Heliocentric radial velocity. (This paper). }  
\tablenotetext{e}{Heliocentric radial velocity. (Brown et al.\ 2008). }  
\tablenotetext{f}{Adopted radial velocity relative to LSR.}  
\end{deluxetable*}

\section{The Radial Velocity Distribution in the $SR$ Field.}

 We begin by discussing the radial velocity distribution of stars in the $SR$ field.
 The $SR$ field covers a pair of overlapping $SERC$~$UKSTU$ objective prism
 fields with plate centers at 03$^{h}$ 40$^{m}$, --65.$^{\circ}$0 
and 03$^{h}$ 48$^{m}$, --60.$^{\circ}$0. Rodgers et al.\ (1993a) list 259 
early-type stars in this field and give radial velocities for 191 of them.
This field has the same galactic coordinates as the 
 (l,b) = (270\arcdeg,--45\arcdeg)
 field observed by Gilmore et al.\ (2002) although the $SR$ field  covers a 
 much larger area to a brighter limiting magnitude. 
  Fig. \ref{Fig1} shows the distributions of the heliocentric radial 
 velocities of both the $SR$ (above) and $NR$ fields; in both cases these   
 histograms refer to early-type stars that are brighter than $V$ = $\sim$15.
 The red histogram in the upper figure shows the distribution of the stars
 with $V$ $\leq$ 18 given by Gilmore et al.\ (2002); the numbers of these 
 turn-off stars have been normalized to be the same as for the early-type 
 stars in the $SR$ field. At this galactic latitude (b = $-45^{\circ}$), the
heliocentric velocity will be about 70\% of the galactic rotational velocity. 
 Gilmore et al.\ describe the velocity distribution for their $V$ $<$ 18 
 stars as that to be expected for a 
 canonical thick-disk with a lag of less than 50 km s$^{-1}$. Their 
 velocity distribution (the red histogram) 
 actually shows a broad peak between 0 and +100 km s$^{-1}$, and so the lag
 seems more like 70 km s$^{-1}$ but this is a minor point. The velocity 
 distribution for the early-type stars in the $SR$ field, on the other hand,
 shows  a broad asymmetrical peak centered on +5 km 
 s$^{-1}$ that continues out to large positive velocities
 (like the red histogram).

 The radial velocity distribution of the early-type stars differs
 quite markedly from that of the turn-off stars:  the distribution for
 the early type stars has a  peak that is close to zero, and so the early type stars
  presumably have a significant old thin disk component (which according to
  Gilmore et al.\ should have a lag of $\sim$20 km s$^{-1}$). 
 ``Early-type stars,'' however, comprise a wide variety of types with
 a range of scale-heights (Preston et al.\ 1994; Knude 1997) and so
 it is probably over-simplistic to characterize them as belonging to
 either thin disk, thick disk or halo as may be done for turn-off stars.

  The early-type 
 stars in the $SR$ field also show a peak in their heliocentric 
 radial velocities at +125 km s$^{-1}$; this corresponds to a lag of +180
 km s$^{-1}$ which is what one  might expect for a prograde halo. Both the 
 early-type stars and the turn-off stars show extended distributions of
 heliocentric radial velocities to
 +350 to +400 km s$^{-1}$ which shows that both contain  a significant
 component of halo stars 
 with retrograde orbits; Gilmore et al.\ (2002) already noted this for their
 turn-off stars that have  $V$ $>$ 18.
   
Too great a correspondence between different {\it tracer} populations should 
 not be expected. Stars as closely related as RR Lyrae stars and BHB stars 
  can show different kinematics in the same field (Kinman et al.\ 2007a), and
 the ratio of 
  BHB to turn-off stars can vary (Bell et al.\ 2010). 
  Nevertheless, further observations of the early-type stars in the
  $SR$ field would be of interest; in particular, an investigation of the
  sharp peak in their radial velocities at $\sim$ +125 km s$^{-1}$. A list of
  these stars and a brief discussion of their possible nature is given in 
  Appendix A.

\section{ The Metallicites and Radial Velocities of the Program Stars in the
    $NR$ Field.}

 \subsection{Metallicities.}

 We now consider the metallicities of the early-type stars in the $NR$
 field.
In their second paper, Rodgers et al.\ (1993b) give the equivalent widths of
the \hbox{Ca\,{\sc ii} K} line and the H$\delta$ Balmer line for 141 of their program 
stars. They do not derive metallicities for the individual stars but, in 
their third paper, Rodgers et al.\ (1993c) used these equivalent widths to show 
that many of their program stars have [Fe/H] $<$--1.0. They conclude that while some
of these metal-poor stars have
 halo kinematics, a significant proportion of those with $V <$14
(which they took to be less than 2.5 kpc above the plane) have kinematics that
are consistent with them belonging to a metal-weak thick disk. 

We derive a metallicity [m/H] for 
each star from a plot of the equivalent width of its 
\hbox{Ca\,{\sc ii} K} line against $(B-V)_{0}$ using the calibration given in Fig.\ 9 of
Clewley et al.\ (2002). These metallicities and the data from which they are
derived are given in Table 11 in Appendix B. The equivalent width  W$_{0}$(K) of 
the \hbox{Ca\,{\sc ii} K} line given by Rodgers et al.\ (1993b) was corrected for an
interstellar component $W(K)$ of 0.6 \AA~.  The value of $W(K)$ is
 discussed in Appendix D where we conclude that a value of 0.3 \AA~ would be 
 more appropriate 
than the 0.6 \AA~correction used by Rogers et al.\ (1993b). We 
therefore also computed metallicities for the program stars using $W(K)$ 
= 0.3 \AA~and these are given in the last column of Table 11 in Appendix B. 
Our derived metallicities [m/H] can only be considered approximate because of 
the combined errors of $W(K)$ and  $(B-V)_{0}$ and the uncertainties in the
 calibration plot. This will be particularly the case 
 for the hottest and most metal-poor stars where the K-line is weak and also 
 for the faintest stars for which the measured equivalent widths are likely to
 be the least accurate. Brown et al.\ (2008) give [Fe/H] for six of these stars. 
The mean difference of the [m/H] in the last column of Table 11 and the [Fe/H] of
Brown et al.\ (2008) for these stars is +0.35$\pm$0.27. We therefore consider
 that although these [m/H] have statistical value, their individual 
values should be taken with considerable reserve.

 \subsection{Radial Velocities of program stars in the $NR$ Field.}

 We now discuss the radial velocities that are available for the early-type
 stars in the $NR$ field.
 Rodgers et al.\ (1993b) give radial velocities for their program stars that
 were obtained with the coud\'{e} spectrograph of the Mt Stromlo 1.88-m telescope.
 The detector was the Mt Stromlo Photon Counting Array which gave a 
 resolution of 2.4~\AA~over the waveband $\lambda\lambda$ 3830 to 4370 \AA.
 The velocities were given with respect to the local standard of rest (LSR) by
 adding +6.0 km s$^{-1}$ to the heliocentric velocities; their probable error was
 estimated to be 18 km s$^{-1}$. Spectra of 18 of these program stars were 
 obtained with RC spectrograph of the Mayall 4-m telescope at Kitt Peak; the
 spectral resolution was 0.8 \AA~over the waveband $\lambda\lambda$ 3880 to 4580
 \AA~(for further details see Kinman et al.\ 1996). The heliocentric radial
 velocities from these spectra were kindly supplied by Dr Nick Suntzeff 
 (priv. comm. 1996, 1998); their estimated errors are $\sim$5 km s$^{-1}$.
 Radial velocities of 8 of the program stars are available from the $Century$
 Survey (Brown et al.\ 2008). These spectra were taken with the FAST 
 spectrograph on the Whipple 1.5-m telescope (resolution 2.3 \AA, waveband
 $\lambda\lambda$ 3450 to 5450, S/N = 30). These velocities have an estimated
 error of 16 km s$^{-1}$. 

 These velocities are compared in Table 4. The mean difference of Stromlo
 $minus$ Kitt Peak is +8.1$\pm$4.9 km s$^{-1}$ where the $rms$ of a single
 difference is 20.3 km s$^{-1}$. The mean difference of Stromlo $minus$ 
 Whipple is --0.7$\pm$8.7 km s$^{-1}$ and the $rms$ of a single difference is
 24.5 km s$^{-1}$. These differences are in satisfactory agreement with 
 quoted errors from the individual sources and there is no evidence for any
 systematic error in the Stromlo velocities. The adopted radial velocities 
 with respect to the LSR are given in the final column of Table 3. Our corrections
 to the LSR have been made separately for each star and {\it not}, as was done
  by Rodgers eta l. (1993b), with a single correction for the whole field.

\begin{deluxetable*}{ccccccccccc}
\tablewidth{0cm}
\tabletypesize{\footnotesize}
\setcounter{table}{5}
\tablecaption{BHB Candidates among Early-Type stars in $NR$}
\tablehead{ 
\colhead{  No.  } &
\colhead{ $V_{0}$       } &
\colhead{ $(B-V)_{0}$ \tablenotemark{a}   } &
\colhead{ $P_{u-B}$  \tablenotemark{b}    } &
\colhead{ $P_{NUV}$  \tablenotemark{c}    } &
\colhead{ $P_{\beta }$  \tablenotemark{d}    } &
\colhead{ $P_{CHSS }$  \tablenotemark{e}    } &
\colhead{ $P_{B2M  }$  \tablenotemark{f}    } &
\colhead{ [m/H]   \tablenotemark{g}    } &
\colhead{ RV   \tablenotemark{h}    } &
\colhead{ Class    \tablenotemark{i} }  \\
   &   &   &   &   &   &   &   &    &km s$^{-1}$ &       \\
}
\startdata
    2 & 12.15&+0.198 &--3  &  0  & --3 &$\cdots$&$\cdots$&--0.8    & +000 &BHB1     \\
    3 &13.15 &--0.031 &--3  &$\cdots$&$\cdots$&$\cdots$&  H  & --2.0   &--218&BHB1    \\
    4 & 13.07&+0.144 & +4  &  +4 &--3&$\cdots$&$\cdots$&--1.2    &     +013 &BHB2     \\
   15 & 14.87&+0.038$\dagger$ & +2  &  +4 &$\cdots$&$\cdots$&$\cdots$& $>$0.0&--211 & BHB3    \\
   17 & 12.33&+0.180 & --3 & +2  &$\cdots$&$\cdots$&$\cdots$& --0.8 & --007 &BHB1    \\
   27 & 12.62&+0.164 & --3 &  0  &$\cdots$&$\cdots$&$\cdots$& --1.5 & +071 &BHB1    \\
   30 & 14.53&--0.049& +4  & +4  &$\cdots$&$\cdots$&$\cdots$& --2.0  &--227 &BHB3\\
   34 & 13.14&+0.237 & --3 & --3 &$\cdots$&$\cdots$&   H & --0.3    &--004  &BHB1   \\
   38 & 12.75&+0.045 & +4  & +4  &  +3   &$\cdots$ &$\cdots$ & --2.2 & --171 &BHB3    \\
   41 & 14.29&+0.023$\dagger$ & +4  & +4  &$\cdots$&$\cdots$  &  H  & --2.2 & --277 & BHB3   \\
   48 & 14.43&+0.020$\dagger$ & +2  &$\cdots$&$\cdots$&$\cdots$ &  H & --0.8 & --121 & BHB1  \\
   52 & 14.06&+0.246 & --3 & --3 &$\cdots$ &$\cdots$ &  H  & --0.7   &--161 &BHB1    \\
   53 & 13.62&+0.122 & +2  &  +4  &$\cdots$&$\cdots$&$\cdots$&  --1.4 &--029 &BHB3 \\ 
   56 & 13.50&--0.261$\dagger$&$\cdots$&$\cdots$&$\cdots$&$\cdots$&   H &$\cdots$&$\cdots$  & BHB1  \\
   60 & 14.62&+0.057 & +2  &  +2  &$\cdots$&$\cdots$ &  H  & --2.6 & --216 &BHB2   \\ 
   63 & 14.60&+0.184 &  0  &$\cdots$&$\cdots$&  0  &$\cdots$ &{\bf--1.13}& --045  & BHB1 \\
   72 & 14.84&+0.282 & --3 & --3 &$\cdots$ &  0  &$\cdots$ &{\bf--0.94}& +002&BHB1  \\
   77 & 11.93&+0.158 & +2  & +2  & --3 &$\cdots$ &  H  & --0.9   &--007 &BHB1    \\
   79 & 13.58&--0.062$\dagger$&  0  & --3 &$\cdots$ &$\cdots$ &  H  & --2.0   &+006  &  BHB1    \\
   83 & 14.63&+0.090 & --4 & --3 &$\cdots$ &$\cdots$ & H:  & --1.2 &--114 & BHB1  \\
   87 & 15.2:&+0.5::$\dagger$ &$\cdots$&--3  &$\cdots$ &$\cdots$ &  H  &$\cdots$ &--281 &  BHB1    \\
   90 & 12.80&--0.063& 0   & --3 &$\cdots$  &--4  &$\cdots$ &{\bf0.00}& --028  &BHB1    \\
   92 & 15.53&+0.037 & +2  &  +2  &$\cdots$&$\cdots$ &$\cdots$ & --1.0 &--228 &BHB2\\
   94 & 13.6:&+0.4:  &$\cdots$ &--3  &$\cdots$ &$\cdots$ &  H  &$\cdots$ &--073  &  BHB1    \\
   99 & 14.53&+0.104 & +4  &  +4  &$\cdots$& +4      &$\cdots$ &{\bf--2.14}&--221  & BHB3 \\ 
  104 & 15.1:&+0.150 &$\cdots$ & +2  &$\cdots$ &$\cdots$ & H &--2.7  &--273 & BHB1    \\
  106 &12.44 &+0.132 &--3     & --3 &$\cdots$ &$\cdots$ & H & +0.0  &+070&BHB1     \\
  110 & 14.91&+0.073 & +4  & +4   &$\cdots$&+4  &$\cdots$ &{\bf--3.00}& --062  &BHB3  \\ 
  115 & 14.37&+0.042 & +4  & +4   &$\cdots$&$\cdots$ &$\cdots$ &  --1.8 &--140  &BHB3  \\ 
  116 & 12.89&+0.195 &--3  & 0   &$\cdots$&$\cdots$&  H  & --1.3   &--017 &      BHB1 \\
  120 & 12.88&+0.198 &  0  &  0   &  +3  &$\cdots$ &$\cdots$ &   --1.6 & --072  &  BHB2 \\ 
  121 & 14.59&+0.124 &--3  &  +4  &$\cdots$ &$\cdots$ &$\cdots$ &   --1.4 &--027 & BHB1  \\
  122 & 14.2:&+0.18:$\dagger$ &$\cdots$ &$\cdots$ &$\cdots$ &$\cdots$ &  M  &$\cdots$&$\cdots$   &   BHB1   \\
  126 & 14.1:&--0.1::&$\cdots$&--3&$\cdots$ &--4 &$\cdots$ & $\cdots$  &{\bf --059}&BHB1  \\
  128 & 12.22&+0.057 & +4  &  +4  &  +3  &$\cdots$ &  H  &   --1.8 & +050  &BHB3 \\
  133 & 14.84&+0.176 & +2  &    0 &$\cdots$&$\cdots$&$\cdots$ & --1.8 &--009 & BHB1     \\
  134 & 13.96&+0.148 & +2  &  +4  &$\cdots$&$\cdots$   &$\cdots$  & --0.4 & --134& BHB3    \\ 
  136 & 11.05&+0.074 & --3 &  +4  &  --3 &$\cdots$  & H   & --1.6 & +023 & BHB1    \\
  138 & 10.78&+0.097 & --3 &$\cdots$ & --3  &$\cdots$ &$\cdots$ & --1.2 &--017 & BHB1    \\
  140 & 13.70&+0.265 &$\cdots$&--3 &$\cdots$ &--4  &   H &{\bf--1.04}& --050  &BHB1 \\
  141 & 14.08&+0.208 & +2  & --3 &$\cdots$&$\cdots$ &$\cdots$ & --1.5 & --125& BHB1  \\
  143 & 11.89&+0.116 & --3  &--3  & --3 &$\cdots$ &   H & --1.4   & +009 &BHB1   \\
  145 & 12.64&+0.027 & +4  & +4   & --3  &$\cdots$ &$\cdots$ & --0.2 & --158  &BHB2   \\ 
  146 & 13.46&+0.058 & +4  & +4   &$\cdots$ &$\cdots$ &  H  & +0.0 & --057 &BHB3  \\
  148 & 14.20&+0.089 &$\cdots$ & 0   &$\cdots$  &  0  &  H  &{\bf--0.25}& --054  & BHB1  \\
  151 & 14.64&+0.009 & +4  & +4  &$\cdots$&$\cdots$ &$\cdots$ &$>$0.0& +039    & BHB3 \\ 
  156 & 13.73&+0.299 &$\cdots$ & --3 &--3  &$\cdots$  &   M &--1.9 &--106 &BHB1    \\
  157 & 12.81&+0.073  & +4  &  +4  &  +4  & +4  &  H  &{\bf--0.75}& --057 &  BHB3\\ 
  158 & 12.47&+0.165  & +4  &  +4  & --3  &$\cdots$ &$\cdots$ &  +0.0 & +032 &BHB2     \\ 
  161 & 12.63&+0.261 &$\cdots$&--3 &$\cdots$ &$\cdots$   &   H &--1.3  & --044 & BHB1     \\
  162 & 13.99&+0.062$\dagger$  &$\cdots$ & +4   &$\cdots$& $\cdots$ &  H  & --2.3 &--025 &BHB2 \\ 
  163 & 14.37&+0.227 &$\cdots$ &--3  &$\cdots$ &  0  &  H  &{\bf--0.91}& +002  & BHB1     \\
  168 & 11.96&+0.158  &--3   &  +2   & --3  &$\cdots$  &  H  & --0.6 & +020 &BHB1  \\ 
  169 & 13.83&+0.047  & +4   & +4   &$\cdots$&$\cdots$ &  M  & --2.0 & --020  &BHB3  \\
\enddata
\tablenotetext{a}{$\dagger$~Outside defining window in $(J-H)_{0}$ $vs.$ $(H-K)_{0}$ plot. See Fig. 10(c) and Fig, 10(d).}
\tablenotetext{b}{BHB class from $(u-B)_{K0}$ $vs.$ $(B-V)_0$ plot. See Sec. 7.5.}
\tablenotetext{c}{BHB class from $(V - NUV)_{0}$ $vs.$ $(B-V)_0$ plot. See Sec. 7.5.}
\tablenotetext{d}{BHB class from Str\"{o}mgren $\beta$ $vs.$ $(B-V)_0$ plot. See Sec. 7.5.}
\tablenotetext{e}{BHB class in $Century$ Survey, See Sec. 7.5.    }
\tablenotetext{f}{BHB class in B2M Survey; See Sec. 7.5.  }
\tablenotetext{g}{Metallicity from Ca\,{\sc ii} K line assuming correction 
 for interstellar component of 0.3\AA~(Rodgers et al.\ 1993b). Metallicities in
 bold-face are from Brown et al.\ (2008).}
\tablenotetext{h}{Adopted radial velocity corrected to LSR. The velocity in bold-face is from Brown et al.\ (2008).}                       
\tablenotetext{i}{ Adopted Class. BHB3 are the most likely to be BHB stars and
 BHB1 are the least likely. 
  } 
                     
\end{deluxetable*}


\section{ The Selection of the Blue Horizontal Branch (BHB) stars in the $NR$ Field.}

\begin{figure}
\includegraphics[width = 3.5in]{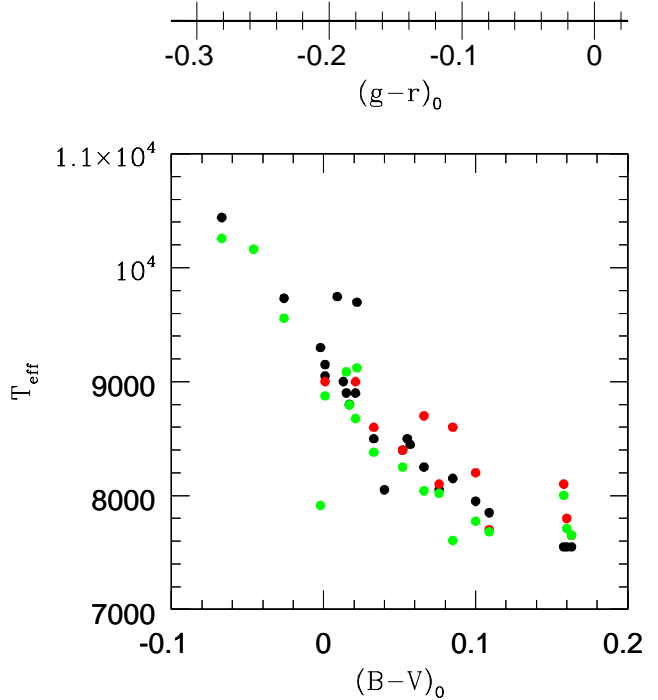}
\caption{ $T_{eff}$ (ordinate) $vs.$ $(B-V)_{0}$ for 25 local BHB stars. The
 $T_{eff}$ are taken from Kinman et al.\ (2000) (black circles), Behr (2003)
 (green circles) and For \& Sneden (2010) (red circles). All $(B-V)_{0}$ are
 taken from Kinman et al.\ (2000). The $(g-r)_0$ scale is approximate.
\label{Fig 4}}
\end{figure}

 The blue horizontal branch of a globular cluster can be quite complex (e.g. as
 in NGC 2808 (Dalessandro et al.\ 2010)) and here we consider only those
 BHB stars with effective temperatures ($T_{eff}$) less than $\sim$10,000 K 
  (called HBA stars by M\"{o}hler (2004)). The primary criterion for selecting
 $field$ BHB stars is that they lie close to
 the Zero Age Horizontal branch (ZAHB) in a plot of $\log$~$T_{eff}$ $vs$
 $\log$ g (see Fig.\ 5 in Behr (2003)). Fig.\ 6 shows a plot of $T_{eff}$ $vs.$
 $(B-V)_{0}$ for 25 local BHB stars; $T_{eff}$ is taken from Kinman et al.\ (2000)
 (black circles), Behr (2003) (green circles) and For and Sneden (2010) (red 
 circles) and the $(B-V)_{0}$ is taken from Kinman et al.\ (2000). This shows that the
 errors are substantial even when high-resolution ($>$ 15,000) and high S/N 
 spectra are available and these limit our ability to distinguish BHB stars 
 from other types --- especially for $T_{eff}$$\geq$10,000~K where the 
 ZAHB and Main Sequence converge in the $T_{eff}$ $vs$ $\log$ g plot. In this 
 paper we use colors as surrogates for $\log$ g and $\log$~$T_{eff}$ and calibrate
 these color-color plots using local bright BHB stars that have been identified
 by the authors referred to above. 

 Preston et al.\ (1991) identified BHB stars by their position in a plot
 of the gravity-sensitive $(U-B)_{0}$ index against $(B-V)_{0}$ over
  the range --0.02$\geq$$(B-V)_{0}$$\geq$0.18. This is an appropriate range of
 $(B-V)_{0}$ since the blue end roughly corresponds to $T_{eff}$ = 10,000 K
 and the red end corresponds to the blue edge of the instability gap 
 Sandage (1990). 
 The Str\"{o}mgren $u$ filter is better than the Johnson $U$ filter for use
 with the Johnson $B$ filter for measuring the Balmer jump. Consequently, in this
 paper we use the $(u-B)_{K0}$ index as defined by Kinman et al.\ (1994) as
 our primary discriminant. This index was measured 
 for 67 stars in the $NR$ field. These include all those stars in 
 the appropriate $(B-V)_{0}$ range with $V_{0}$ $\leq$ 15.0 except for stars
 97, 104, 122, 126, 139, 148 and 162. For these latter stars we used the 
  $(NUV-V)_{0}$ index as a discriminant as
 described by Kinman et al.\ (2007b).

\begin{figure}
\includegraphics[width = 3.5in]{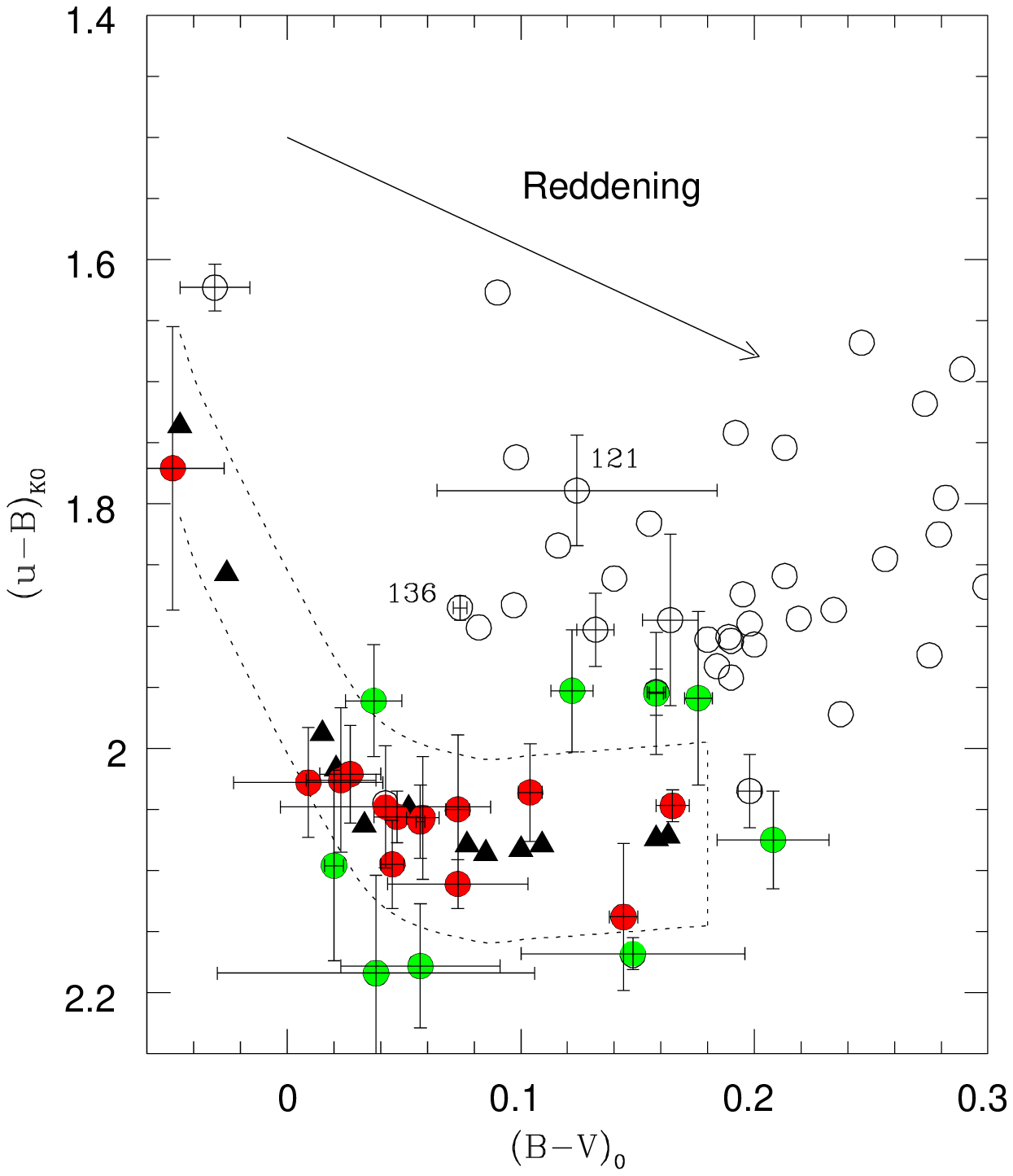}
\caption{ The ordinate $(u-B)_{K0}$ is defined in Kinman et al.\ (1994). 
 The abscissa is Johnson $(B-V)_{0}$. The data for the local BHB stars
 (filled triangles) are taken from Kinman et al.\ (2000). The location of these
 local BHB stars is used to define the window shown by dotted lines. 
 The BHB star candidates are shown by red filled circles if they lie within 
 this window and by green filled circles if their error bars lie within the
 window. The other program stars are shown by open circles. 
\label{Fig 5}}
\end{figure}

\subsection{Selection using the $(B-V)_{0}$ $vs.$ $(u-B)_{K0}$ Plot.}

 The $(B-V)_{0}$ $vs.$ $(u-B)_{K0}$ plot  is shown 
  in Fig.\ 7. Here, nearby BHB stars whose classification
 is secure (Kinman et al.\ 2000; Behr 2003) are shown as filled triangles that
 lie on a well-defined curve. The dotted lines in Fig.\ 7 enclose an area
in which $(B-V)_{0}$ $\leq$ 0.18 and  $(u-B)_{K0}$ is within $\pm$0.075 mag. of 
the curve defined by the nearby BHB stars. Stars that lie within this area are 
 shown as red filled circles while those
whose error bars lie within or very close to this area are shown as green filled
circles. Error bars are also shown for the five stars (3, 27, 106, 121, 
 \& 136) for which there is some evidence that they might be BHB stars.

\begin{figure}
\includegraphics[width = 3.5in]{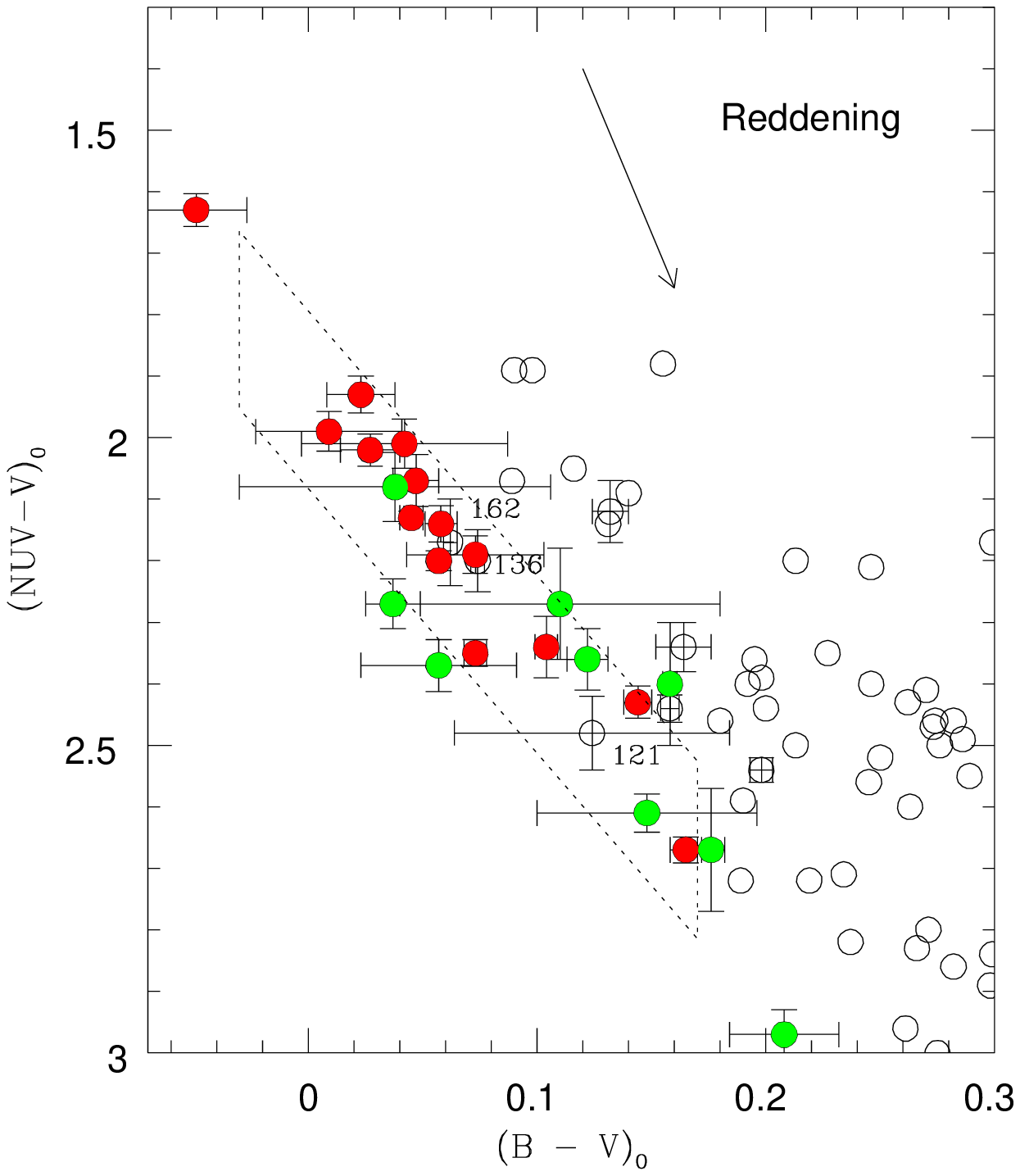}
\caption{ The ordinate is the de-reddened difference between the $GALEX$ $NUV$
 magnitude (effective wavelength 2267\AA) and the Johnson $V$ magnitude. 
 The abscissa is the Johnson $(B-V)_{0}$ color. The dotted parallelogram is
 the expected location of the BHB star candidates (Kinman et al.\ 2007b). The 
 symbols have the same meaning as in (Fig 7) and refer to their location in
 the $(u-B)_{K0}$ $vs.$ $(B-V)_{0}$ plot (Fig. 7).
\label{Fig 6}}
\end{figure}

\subsection{Selection using the $(B-V)_{0}$ $vs.$ $(NUV - V)_{0}$ plot.}

The $(B-V)_{0}$ $vs.$ $(NUV - V)_{0}$ plot is shown in Fig.\ 8. The dotted 
parallelogram (taken from Kinman et al.\ 2007b) shows the expected location of the
BHB stars. The red and green filled circles show the stars that are 
in or close to the BHB location in Fig.\ 7;  these stars also lie
satisfactorily close to the defining area in Fig.\ 8. There are, however, three
stars (121, 136 \&  162) that are BHB stars according to their $(NUV-V)_{0}$ 
index although stars 121 and 136 are not BHB stars according to their $(u-B)_{K0}$ index.
The $NUV$ magnitude of star 136 is close to the saturation limit for this $GALEX$
magnitude (Morrissey et al.\ 2007) so its $NUV$ error may have been 
 underestimated. Its identification as a BHB star is therefore uncertain.

\begin{figure}
\includegraphics[width = 3.5in]{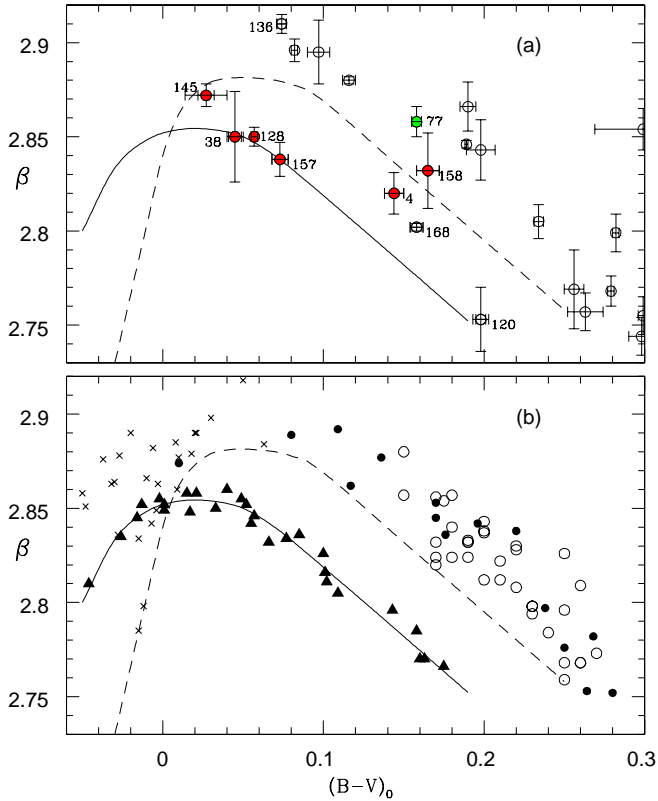}
\caption{ In both plots the ordinate is Str\"{o}mgren $\beta$ and the abscissa
 is Johnson $(B-V)_{0}$. In the lower figure 9(b), the local BHB stars are
shown by filled triangles and their location is indicated by the solid curve. 
The data for these stars are taken from Kinman et al.\ (2000). The other symbols
show the locations of nearby non-BHB stars and the lower limit of $\beta$ for
these stars is shown by the dashed line. The data for these non-BHB stars come 
from Gray \& Garrison (1989) (open circles), Stetson (1991) (filled
circles) and Crawford et al.\ (1972) (crosses). The solid and dashed curves
are repeated in Fig. 9(a) which shows the locations of the program stars.
The BHB candidates 
derived from the $(u-B)_{K0}$ $vs.$ $(B-V)_{0}$ plot (Fig. 7) are shown by
red and green filled circles.  The other program stars are
shown by open circles. 
\label{Fig 7}}
\end{figure}

\begin{figure}
\includegraphics[width = 3.5in,angle=0]{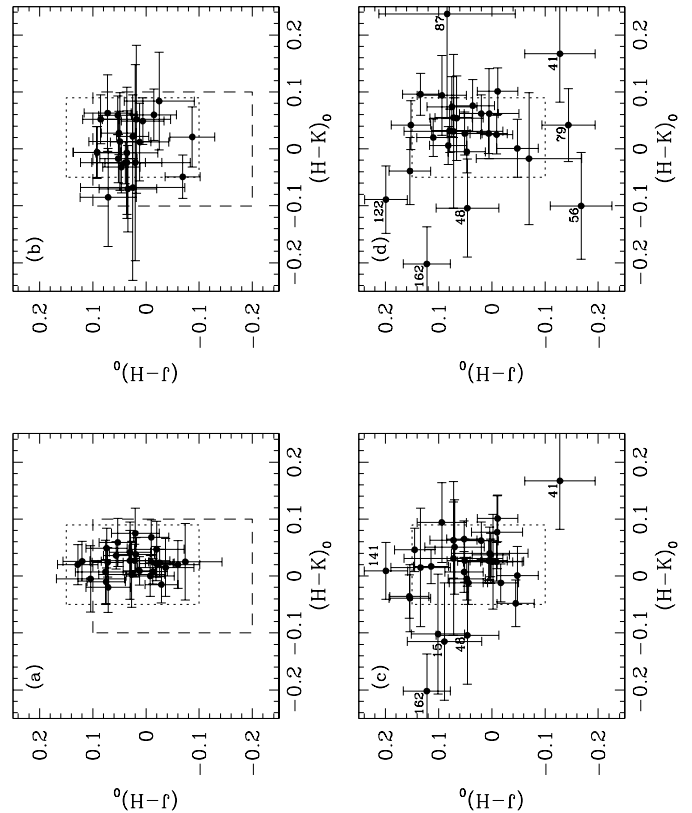}
\caption{ The 2MASS color $(J-H)_{0}$ (ordinate) $vs.$ $(H-K)_{0}$ for 
     (a) local BHB stars; (b) BHB stars from the $CHSS$ sample (c) BHB
   candidates from $NR$ field; (d) BHB candidates classified H and M in the
   B2M survey that are in the $NR$ field.  
   The significance of the dashed and dotted windows 
   is given in the text. Stars that lie outside these windows are numbered.
\label{Fig 8}}
\end{figure}

\subsection{Selection using the Str\"{o}mgren $\beta$ $vs.$ $(B-V)_{0}$ Plot.}

The Balmer lines of BHB stars are narrower than those of higher-gravity
main sequence stars of the same effective temperature and this property has been used
to identify BHB stars (Searle \& Rodgers, 1966; Pier, 1983). The Str\"{o}mgren 
$\beta$ index can be used as a surrogate for the commonly used D$_{0.2}$ width as
 shown in the Str\"{o}mgren $\beta$ $vs.$ $(B-V)_{0}$ plot (Fig.\ 9b) in which 
 the nearby BHB stars (filled circles) with +0.03$\leq$$(B-V)_{0}$$\leq$+0.18 
have a different location from the non-BHB stars. The data for the
 non-BHB stars in Fig 9(b) were taken from Crawford et al.\ (1972) (crosses),
Gray \& Garrison (1989) (open triangles) and Stetson (1991) (inverted filled
triangles) while those for the nearby BHB stars were taken from Kinman et al.\ 
 (2000). Fig.\ 9(a) shows the same plot for the brighter of our program stars 
for which $\beta$ could be measured accurately enough with a 0.9-m telescope. 
  We use the same symbols as in Fig.\ 7. Four of this sample 
  (nos.\ 38, 120, 128 \& 157) lie on the
  line that defines the BHB stars --- supporting this classification
for them. Five (nos.\ 4, 77, 145, 158 and 168) lie well above the BHB line.
  This casts doubt on their classification as BHB stars and we 
 discuss this further in Sec. 8.2. Star 136, whose classification as a BHB star
 depends upon a questionable $NUV$ magnitude, lies well above the BHB line (in
 (Fig. 9(b)) showing that its H$\beta$ is  too broad for
 it to be a BHB star. 

\subsection{Selection using the  $(J-H)_{0}$ $vs.$ $(H-K)_{0}$ Plot.}

 Brown et al.\ (2008) used the window 
 --0.20 $<$ $(J-H)_{0}$ $<$ 0.10 and --0.10 $<$ $(H-K)_{0}$ $<$ 0.10 in making 
 their initial selection of 
 stars for the Century Survey. Fig.\ 10(a) shows the position of the local
BHB stars (Kinman et al.\ 2000) in this diagram. These stars are
concentrated in an area (outlined by the dotted rectangle) that is significantly
 smaller than the $CHSS$ window (outlined by the dashed rectangle). 
 Fig.\ 10(b) shows the same diagram for a sample of 24 BHB stars from the $CHSS$
survey that have 80$^{\circ}$ $\leq$ l $\leq$ 100$^{\circ}$ and 
--35$^{\circ}$ $\geq$ b $\geq$ --55$^{\circ}$; these stars are also contained by the
 dotted rectangle (--0.10 $<$ $(J-H)_{0}$ $<$ 0.15 and 
  --0.05 $<$ $(H-K)_{0}$ $<$ 0.09).
 This confirms that this smaller window is adequate for selecting BHB stars.
 Fig.\ 10(c) 
 shows the position of our $NR$ survey BHB candidates in this diagram. Two stars,
 41 and 162 have  $JHK$ colors that lie outside the window and consequently are
 hotter than the other BHB candidates.

\subsection{The Selection of BHB Stars from the Candidates.}

  We make our final selection from the fifty four BHB
  candidates that are given in Table~5 where a weight is given to
  each star for each selection method according to the 
  probability of its being a BHB star. For selection using 
  $(u-B)_{K0}$ and $(NUV - V)_{0}$, stars in the defining box were given 
  weight +4; those whose error bars intersected the defining box were given 
  weight +2. Others were given weights 0 and --3 according to their distance
  from the defining box. For selection using Str\"{o}mgren $\beta$, 
  stars on the defining line were given weight +3; others were given weight
  --3. For stars classified by the CHSS survey, those classified as BHB stars
  were given weight +4, the unclassified were given weight 0 and those with
  specific non-BHB classifications were given weight --4. We also included 
  the 26 stars in the $NR$ field that
  the $B2M$ survey classifies as having a high (H) or medium (M) probability
  of being a BHB star. Weight 0 was given for this classification.

  The weights from each selection method were added to give a total weight
  (W) for each star.
  The fourteen stars with 
  W $\geq$ +6 are classified BHB3 and are likely BHB stars.
  The seven stars with $2 < W < 6$ are classified 
  BHB2 and are possible BHB stars.
  The remaining thirty three stars with W $\leq$ +2 are classified BHB1 and 
  are unlikely to be BHB stars.  The mean properties of the stars in the 
  three classes are given in Table~6.

\begin{figure}
\centerline{\includegraphics[width = 2.75in]{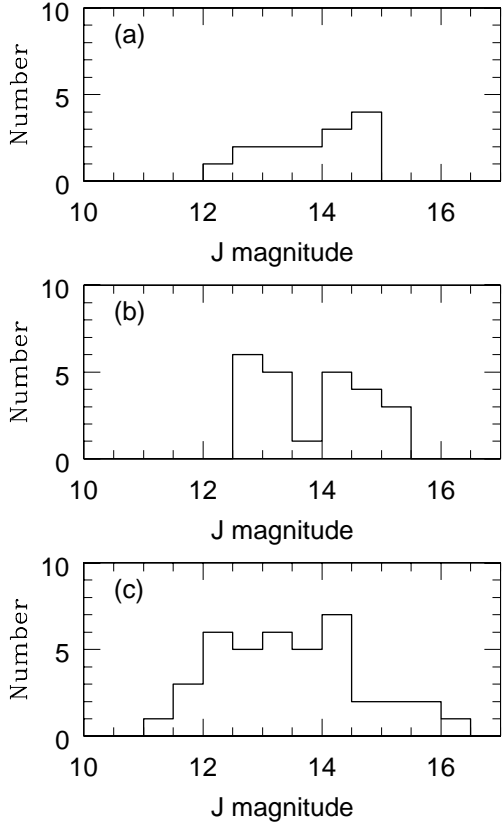}}
\caption{ The distribution with 2MASS J magnitude of 
    (a) The BHB stars classified as BHB3 in Table 5 (70 deg$^{2}$).
    (b) Nearby sample of  BHB stars in the $CHSS$ survey (119 deg$^{2}$).
    (c) Nearby sample of  stars classified as ``H ''and ``H:'' (high probability of
    being  BHB stars in the $B2M$ survey (66 deg$^{2}$).
\label{Fig 9}}
\end{figure}

\begin{deluxetable*}{cccccccccc}
\tablewidth{0cm}
\tabletypesize{\footnotesize}
\setcounter{table}{6}
\tablecaption{ Properties of the classes of BHB candidates.}                        
\tablehead{ 
\colhead{  Class \tablenotemark{a}  } &
\colhead{  N  \tablenotemark{b}  } &
\colhead{ $<V_{0}>$ \tablenotemark{c}  } &
\colhead{Range  \tablenotemark{d}} &
\colhead{ $<(B-V)_{0}>$ \tablenotemark{e}  } &
\colhead{Range \tablenotemark{f}} &
\colhead{ $<[m/H]>$ \tablenotemark{g}  } &
\colhead{Range \tablenotemark{h}} &
\colhead{$<$RV$_{lsr}$$>$\tablenotemark{i}   } &
\colhead{Range \tablenotemark{j}} \\
   &  & &in $V_{0}$ & &in $(B-V)_{0}$ &  &in [m/H] &km s$^{-1}$&in $RV_{lsr}$ \\
}

\startdata
 BHB3 & 14& 13.91& 2.69& +0.057& 0.198&--1.41&3.00& --108$\pm$29&327   \\
 BHB2 & 07& 13.60& 3.06& +0.099& 0.141&--1.27&2.60& --93$\pm$44&260   \\
 BHB1 & 33& 13.43& 4.42& +0.148& 0.761&--1.13&2.70& --53$\pm$16&352   \\
\enddata

\tablenotetext{a}{ BHB Class as defined in Sec. 7.5 }
\tablenotetext{b}{ Number of stars in class. }
\tablenotetext{c}{ Mean $V_{0}$ for stars in class. }
\tablenotetext{d}{ Range in $V_{0}$ for stars in class. }
\tablenotetext{e}{ Mean $(B-V)_{0}$ for stars in class. }
\tablenotetext{f}{ Range in $(B-V)_{0}$ for stars in class. }
\tablenotetext{g}{ Mean [m/H] for stars in class. }
\tablenotetext{h}{ Range in [m/H] for stars in class. }
\tablenotetext{i}{ Mean radial velocity corrected to LSR for stars in class. }
\tablenotetext{j}{ Range in radial velocity corrected to LSR for stars in class. }

\end{deluxetable*}

\begin{deluxetable}{cccccc}
\tablewidth{0cm}
\tabletypesize{\footnotesize}
\setcounter{table}{7}
\tablecaption{Mean Galactocentric Radial Velocities and their Dispersions for 
 $CHSS$  and $NR$ BHB stars with galactic longitudes near 90$^{\circ}$.} 
\tablehead{ 
\colhead{Z range\tablenotemark{a}} & 
\colhead{$\langle$Z$\rangle$ \tablenotemark{b}} & 
\colhead{n \tablenotemark{c}} & 
\colhead{Area } & 
\colhead{$V_{gal}$ \tablenotemark{d}} & 
\colhead{$\sigma$ \tablenotemark{e}} \\
          kpc & kpc &    & Deg$^{2}$ &km s$^{-1}$  & km s$^{-1}$         \\
}

\startdata
      $<$ 4 & 2.68 &  36 &281&  +34.9$\pm$15.9 & 94.2$\pm$11.1  \\
             &      &     &   &             &                \\
      $>$ 4 & 5.78 &  25 &281&  +12.1$\pm$27.0 &132.1$\pm$18.7  \\
\enddata

\tablenotetext{a}{Range in Z.}  
\tablenotetext{b}{Mean value of Z.}  
\tablenotetext{c}{Number of BHB stars in Field. }
\tablenotetext{d}{Mean Galactocentric Velocity with error. }  
\tablenotetext{e}{Dispersion in Galactocentric Velocity with error. }  
\end{deluxetable}

\section{Discussion.}

   We conclude that the 14 stars that we classifed as BHB3 (Sec.\ 7.5) 
   have a high probability of being BHB stars and that the 33 stars 
  classified as BHB1 are unlikely to be BHB stars. 
  BHB2 is an intermediate class whose nature needs clarification. The mean
  properties of these classes are given in Table 6; the stars in class 
  BHB3 (the BHB stars)
  are fainter, bluer, more metal-poor and have a more negative radial 
  velocity ($RV_{lsr}$) than those in class BHB1 (the non-BHB stars). The 
  ranges in these parameters are also smaller for the BHB stars.
  The Shapiro-Wilk test (Shapiro \& Wilk, 1965) shows that 
  the $RV_{lsr}$ distribution of the BHB3 stars shows no departure from normality
  as we would expect for a kinematically homogeneous group, 
  The $RV_{lsr}$ distribution of those in BHB1, however, shows a very significant 
  departure from normality. Evidently, the BHB1 class (non-BHB stars)
  contain both disk and halo stars 
  (c.f. Fig 8.\ in Brown et al.\ (2008)). 

  The data in color-color plots are affected both by errors in the 
  correction for interstellar extinction and also systematic errors in the 
  photometric systems. It is not easy to evaluate such errors and so it is
  important to compare our results with those from other surveys that depend
  on other selection methods.

\subsection{Comparison with $CHSS$ and $B2M$ surveys.}

  The $CHSS$ survey has 10 stars that are in the $NR$ survey area. They classify
  3 of them as BHB stars in agreement with our classification as BHB3.
  They either do not classify the remaining 7 or classify 
  them as a type other than BHB; we call all these 7 class BHB 1.
  The $CHSS$ survey covers an area of 119 deg$^{2}$ 
  in the sky between 80$^\circ$ $\leq$ l $\leq$ 100$^{\circ}$ and 
  between --55$^\circ$ $\leq$ b $\leq$ --35$^{\circ}$. In this field they 
  classify 24 stars as type BHB and 6 as BHB/A. The corresponding surface 
  densities are 0.20$\pm$0.04 and 0.05$\pm$0.02 stars deg$^{-2}$ 
  respectively 
  for the BHB and BHB/A types. In the $NR$ field, the density of the 
  class BHB3 stars is 0.19$\pm$0.05 stars deg$^{-2}$ while that 
  of type BHB2 is 0.09$\pm$0.04 stars deg$^{-2}$. Although the $CHSS$ survey 
  extends to slightly fainter magnitudes (Fig.\ 11b), the $CHSS$ BHB stars
  and our BHB3 stars 
  have a similar magnitude range.
   There is therefore good agreement overall between the $CHSS$ and our 
  present survey. 

  The $B2M$ survey has 26 stars that are classified H or H: and 3 as M 
  that are in the $NR$ survey area.
   Only 5 of these (4 class H and 1 class M) are in our class BHB3 
   and so are likely to be BHB stars while 17 (15 as H and H: and 2 as M) 
  are in class  BHB1 and so are unlikely to be BHB stars. To get a 
  larger sample of the $B2M$ survey in this part of the sky, we took those  
  in the R.A. range 23:00 to 23:40 and Declination range +02$^{\circ}$ to 
  +20$^{\circ}$. Forty $B2M$  stars in this region are classified as either
  H or H: and have a mean galactic longitude (l) of 89$^{\circ}$ and mean 
  galactic latitude (b) of --47$^{\circ}$; they
  are contained in an area of 66.2 deg$^{2}$. This corresponds to 
  0.60$\pm$0.10 stars deg$^{-2}$ or three times the surface density of the 
  BHB stars found by the $CHSS$ and our survey. Some this excess may be caused 
  by the $B2M$ survey having a slightly deeper limiting magnitude but it
  is mainly caused by the inclusion of non-BHB stars.  
  The  $B2M$ sample (Fig.\ 11(c))
  contains many stars that are significantly brighter than those in the other
  two surveys. We conclude that only about a third to one
  half of the stars that $B2M$ classify as having a high probability of being 
  BHB stars are actually BHB stars. This result is consistent with that of
  Ortiz et al.\ (2010) who made a spectroscopic survey of 43 $B2M$ stars. They
  found only 13 stars (30\% of their sample) could be reliably identified as
  BHB stars, 25\% could be identified as some other type and for 40\% the type
  was uncertain.  Many of the stars in their
  sample had $(B-V)_{0}$ $\leq$ 0.0 or $T_{eff}$ $\geq$ 10,000$^{\circ}$ and the
  classification of such stars is difficult (Behr 2003).

  In the Appendix C, we take larger samples of the $CHSS$ and $B2M$ BHB stars 
  and examine the probabilities that they are BHB stars given by Smith et al.\  
  (2010). Many of the stars that $B2M$ classify as having a high probability of
  being BHB stars are not in the Smith et al.\ catalog and are  likely 
  to be too blue for the $B2M$  classification to be correct.

\begin{figure}
\includegraphics[width = 3.5in]{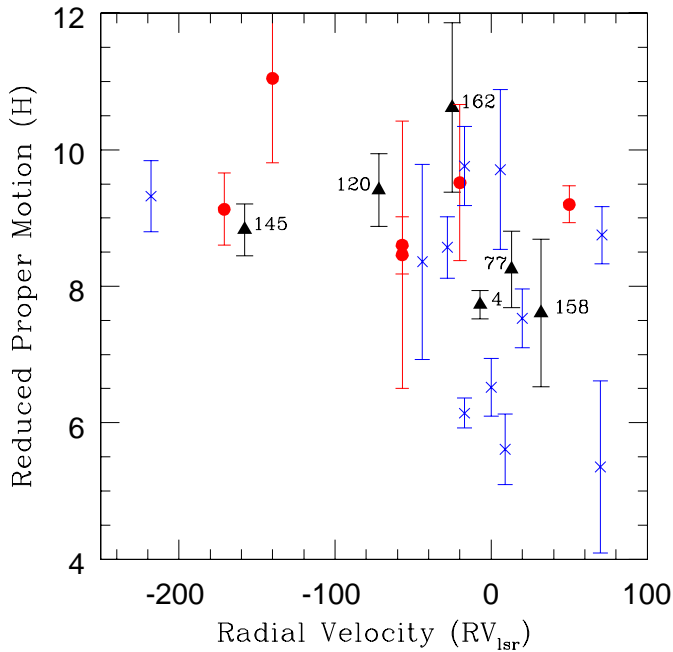}
\caption{ The reduced proper motion H (ordinate) $vs.$ radial velocity
    in km s$^{-1}$ ($RV_{lsr}$ for stars of (a) class BHB3 (red filled
    circles), (b) class BHB2 (black filled triangles) and (c) class BHB1 
    (blue crosses). 
\label{Fig 12}}
\end{figure}

\subsection{ The  Ambiguous class in the present sample.}

  The ambiguous class (BHB2) contains stars of two types. 
  The first type (stars 60,
   92 and 162) are relatively faint ($V$ $>$ 14) stars for which the 
  observations have too low a weight to assign them to class BHB3.
  Stars 60 and 92 have large negative $RV_{lsr}$ and are likely to be halo
  stars, while star 162 has a low $RV_{lsr}$ and a location on the $(J-H)_{0}$
  $vs.$ $(H-K)_{0}$ that differs from that of other BHB stars (Fig.\ 10(c)).
 The second type (stars 4, 120, 145, and 158 in BHB2 and 77 in BHB1) are those
  where the classification based on Str\"{o}mgren $\beta$ differs from that
  based on $(u-B)_{K0}$ and/or $(NUV - V)_{0}$. These latter are brighter
  ($V$ $<$ 13.2) and, except for star 145, have lowish  $RV_{lsr}$ and 
  ${(B-V})_{0}$ $\geq$ 0.14. 

  The reduced proper motion $H$ is defined (following Stetson 1981) by:
 \begin{equation}  
      H  =  V_{0} + 5 + 5\log\mu  
 \end{equation}  
    where $V_{0}$ is the extinction-corrected $V$ magnitude, and $\mu$ is the
 {\it total} proper motion in arcsec per year. Fig.\ 12 is a plot of H against
 the radial velocity of the 23 stars for which significant proper motions are
 available from the UCAC3 catalog (Zacharias 2010). The BHB stars (red filled
 circles) have a wide range of radial velocity but a restricted range of 
 reduced proper motion (H). A few of the non-BHB stars (blue crosses) are also 
 found in the same location as the BHB stars but most have a broad range in
 H and a smaller $V_{lsr}$ than the BHB stars. The ambiguous BHB2 class 
 (black triangles) show an intermediate location and so their kinematics are
 as ambiguous as their other properties. We judge that among these stars, 
 60, 92, 120 and 145 are the most likely and 
 4, 77, 158 and 162 are the least likely to be BHB stars. 
 Stars 60, 90, 120 and 145 
  have a mean $RV_{lsr}$ of 
 --118 km s$^{-1}$ and so their addition to our BHB3 sample of BHB
 stars would only change the mean velocity to --111$\pm$24 km s$^{-1}$. The
 surface density of these 4 stars is 0.06$\pm$0.03 deg$^{-2}$ which is 
 similar to the surface density of 0.05$\pm$0.02 deg$^{-2}$ for the BHB/A 
 types of the CHSS survey.  

 {\it We conclude that both the $CHSS$ and our present survey are of similar 
 quality. The selection criteria in both ensure that the stars which they 
 classify as BHB stars have few interlopers. Both surveys, however, identify
 significant numbers of stars ($\sim$25\% as many as in the BHB class) that
 have perhaps a 50\% chance of being BHB stars. The sum of the BHB and BHB/A
 classes will therefore be more complete but $\sim$10\% of its content will be
 misidentified.}

\subsection{Disk or Inner Halo BHB stars?}

\subsubsection{Radial Velocities.}
  Radial velocity alone can be used to discriminate the population type at
  galactic longitudes 90$^{\circ}$ and 270$^{\circ}$, and we consider the BHB
  stars in several fields with these galactic longitudes. The first field 
      contains the 26 BHB stars in the $CHSS$ with
 80$^{\circ}$ $<$ l $<$ 100$^{\circ}$ and +55$^{\circ}$ $<$ b $<$ +35$^{\circ}$.
  The second field  contains the 24 BHB stars in the $CHSS$ with 
 80$^{\circ}$ $<$ l $<$ 100$^{\circ}$ and $-55^{\circ}$ $<$ b $<$ $-35^{\circ}$.
  The third field is the $NR$ field that contains 14 BHB stars. Taking into 
  account overlapping, there are 61 BHB stars in these fields;
  their mean galactocentric radial velocities and their dispersions and other
  properties are given in Table 7.

 If a significant fraction of the BHB stars belong to the disk, we expect that
 BHB stars with Z $<$ 4 kpc will have more positive galactocentric radial velocities and
 smaller radial velocity dispersions than those with Z$>$ 4 kpc. If such an
 effect is present in the sample in Table 7, it is at the limit of
 significance.  Disk BHB stars should not have a normal radial velocity disitribution.
 Application of the Shapiro-Wilks test to the 36 BHB stars in the Z $<$ 4 kpc
 and the 17 BHB stars with Z $<$ 2.5 kpc in our sample give W = 0.951 and 0.923
 respectively; neither sample shows a significant departure from a normal 
 distribution.

 The evidence from the radial velocities therefore suggests
 that any disk component of the BHB stars must be quite small; this agrees with the
previous analysis of the proper motions of $CHSS$ BHB stars by Kinman et al.\
(2009).  For stars with Z $<$ 4 kpc, there is some evidence that the BHB stars
 show slightly prograde rotation (c.f. Carollo et al.\ 2007\footnote{ For a
 criticism of this paper, however, see Sch\"{o}nrich et al.\ (2010).}).
  but for Z$>$4, 
 there is no evidence for rotation; this agrees with the analysis of the 
 motions of 1700 subdwarfs in the contiguous Stripe 82 field (Smith et al.\
 2009).

\begin{figure}
\includegraphics[width = 3.5in]{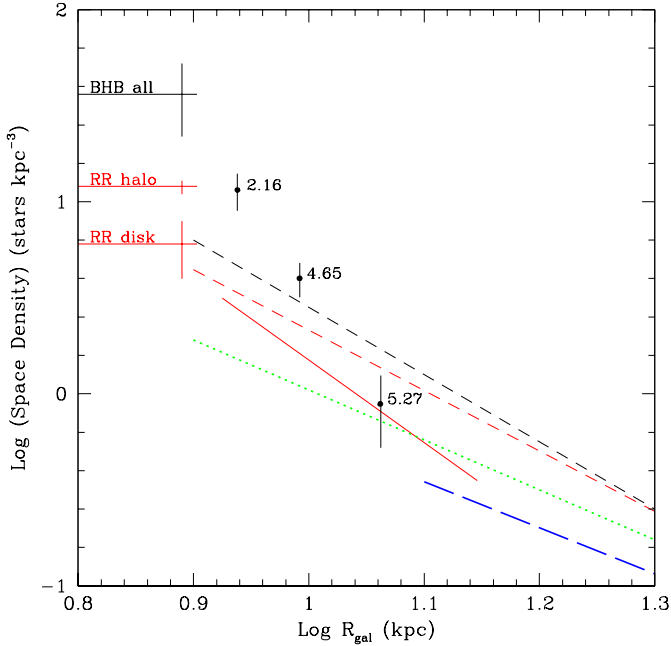}
\caption{ The ordinate is the Log of the space density in stars kpc$^{-3}$
    and the abscissa is the Log of the Galactocentric distance in kpc. The
 black filled circles refer to the BHB stars 
 described in Table 7 and the numbers beside 
 them give their mean height $\langle$Z$\rangle$ from the Galactic 
plane (in kpc). The black dashed line is for a spherical halo of BHB stars  
with Z $>$ 5 kpc and is taken from Kinman et al.\ (1994). 
    The green dotted line shows the extrapolation of the BHB space densities
    for the Northern field of De Propris et al.\ (2010). The blue line 
    (long dashes) shows the BHB space densities of their Southern field. 
The red solid line shows the $QUEST$ RR$ab$ Lyrae space densities (Vivas \&
   Zinn, 2006) and the red dashed line shows the $LONEOS$ RR$ab$ Lyrae space
    densities (Miceli et al.\ 2008). The local space densities of the 
     BHB and RR Lyrae stars are from Kinman et al.\ (2009). 
\label{Fig 13}}
\end{figure}

\subsubsection{Space Densities.}

 The space densities of halo stars are generally represented in terms of a 
 power-law of 
 the galactocentric distance (R$_{gal}$) and a flattening which may either be
 constant or a function of R$_{gal}$ such as that introduced by Preston et 
 al.\ (1991). Recent estimates of the power-law exponent and the flattening
 (c/a)  vary quite widely (see Table 2 in Miceli et al.\ (2008)). 
 The older studies tended to find steeper exponents such as the --3.2 and
 --3.5 for RR Lyrae stars and BHB stars respectively by Preston et al.\ (1991), 
 the --3.5 found for BHB stars
 with Z $>$ 5 kpc by Kinman et al.\ (1994) and the --3.53 for
 RR Lyrae stars by Wetterer and McGraw (1996). 
More recently Miceli et al.\ used their $LONEOS$ RR Lyrae survey to
 model a spherical halo with exponents of --2.26$\pm$0.07 for Oosterhoff Type
 I RR Lyrae stars and --2.88$\pm$0.11 for those of Oosterhoff type II. They
 also found a model with variable flattening for both Oosterhoff types with 
 an exponent of --3.15$\pm$0.07. A similar exponent (--3.1) was found 
  by Vivas and Zinn (2006) from their 
 $QUEST$ RR Lyrae survey using a model with variable flattening that 
 excludes  regions that have pronounced overdensities. We consider the
 $LONEOS$ and $QUEST$ surveys (which refer only to type $ab$ RR Lyrae
 stars) to be the most reliable now available. The $LONEOS$ space densities
 have been recomputed for an RR Lyrae M$_{V}$ of +0.55 (as used by Vivas
 \& Zinn) and are shown by red dashed lines in Fig.\ 13. The $QUEST$ space
 densities are shown by a full red line in this figure. Both surveys
 predict a space density of 5 RR$ab$ stars kpc$^{-3}$ at the solar R$_{gal}$
of 8.0 kpc. This is appreciably less than the observed local density for 
 both disk and halo 
 type $ab$ RR Lyraes (14$\pm$2 stars kpc$^{-3}$) which itself is likely
 to be a lower limit. About a third of these local stars are the disk
 RR Lyrae stars that are more metal-rich than the halo RR Lyraes and
 which have different kinematics, but there are still about twice as many
 local $halo$ type $ab$ RR Lyrae stars as would be predicted by the $LONEOS$
 and $QUEST$ surveys \footnote{We noted in Sec. 4.2 that the surface   
 densities of the RR Lyrae stars that we find in the $NR$ field 
  roughly agree with those found in the $LONEOS$  
 survey if we take into account that only the higher amplitude variables 
 are found in the     $ASAS-3$ and $NSVS$ surveys.}.

   We calculate space densities ($\rho$) in stars kpc$^{-3}$ for the BHB
   stars near galactic longitude 90$^{\circ}$ 
    (Table 7) for three volume elements bounded by distances
   of 0.0, 4.0, 
   6.5 and 8.5 kpc in which the survey is thought to be complete. These are
  shown in a plot of $\log \rho$ $vs.$ $\log$ R$_{gal}$ in Fig.\ 13 where the
   mean height above the plane ($\mid$Z$\mid$) of each volume element is
   shown next to the corresponding space density. Also plotted are the space
   densities of the local BHB and RR Lyrae stars (Kinman et al.\ 2009). Fig.
   13  may be compared with a similar plot (Fig.\ 15 in Kinman et al.\ 1994)
   where both BHB stars and RR Lyraes have comparable space densities that 
   (for Z $>$ 5 kpc) could be represented by a spherical halo with a power
   law exponent of --3.5 (shown by the black dashed line in Fig.\ 13). 
    De Propris et al.\ (2010) found exponents of --2.6 and --2.4 for
   spherical halo fits to the Northern and Southern fields of their BHB 
   survey (shown in Fig.\ 13 by green dotted lines and long blue dashed lines 
   for their Northern and Southern fields respectively). Their Northern field
   lies on the celestial equator with an R.A. between 09$^{h}$ 30$^{m}$ and
   14$^{h}$ 30$^{m}$
   and their  Southern field covers the South Galactic Pole.
    The outermost point ($\mid$Z$\mid$ = 5.27 kpc) for our sample of BHB stars
   near galactic longitude 90$^{\circ}$ is in good agreement with an 
 {\it extrapolation} of the power-law which they fit to the space densities of
  their Northern field. 
  Our two innermost points for $\mid$Z$\mid$ = 2.16
   and 4.65, however, are much larger than the predictions of
  these models and show that $\rho$ increases towards the galactic plane 
  more steeply than they predict from analyses of  stars in the outer
  halo;  this is  also shown by Kinman et al.\ (1994) in their Fig.\ 15. 

   {\it  We know, however, both from the 
  kinematics of local BHB stars, from our present analysis of the radial
  velocities of BHB stars with galactic longitudes near 90$^{\circ}$, 
   and from the proper motion analysis of $CHSS$ stars 
  at the North Galactic Pole (Kinman et al.\ 2009) that this
  excess of BHB stars near the plane has a disk-like spatial distribution but is 
   largely composed of stars that have zero or low galactic rotation. 
  We conclude that the BHB stars are revealing a spatially flattened, non-rotating 
  ``inner halo'' as described  by Kinman et al.\ (2009) and Morrison et al.\  (2009).}

\section{Conclusion.}

  The distribution of heliocentric radial velocites $V_{hel}$ of the
  early-type stars in the $SR$ field of Rodgers et al.\ (1993a) 
  (l = 270$^{\circ}$, b = --45$^{\circ}$) does not show any evidence for the
  anomalous disk  of faint turn-off stars found by Gilmore et al.\ (2002) 
  in the same galactic location. 
  There is a peak in these $SR$ velocities, however, that corresponds to
  the slightly prograde galactocentric radial velocity of --30 km s$^{-1}$.
  These stars are listed and discussed in Appendix A; it is unlikely that
  more than a third of these stars could be BHB stars. There is little
  evidence for a corresponding peak in the radial velocities in the $NR$
  field. The mean galactocentric radial velocity of the BHB stars near 
  galactic longitude (l) = 90$^{\circ}$ that have 
  Z$<$ 4 kpc  (Table 7) may be slightly prograde but the 
  Shapiro-Wilk test shows that this group has a normal distribution and so
  is likely to consist of a homogeneous group of halo stars rather than a  
  mix of disk and halo stars. The BHB stars in Table 7 
  with Z$>$ 4 kpc show no rotation.

  New photometry and supplementary data for 
  the early-type stars in the $NR$ field are given in Table 1 of this paper.
 Plots of 
   $(u-B)_{K0}$ $vs.$ $(B-V)_{0}$,
  ($NUV - V)_{0}$ $vs.$ $(B-V)_{0}$,  and  
 Str\"{o}mgren $\beta$ $vs.$ $(B-V)_{0}$ are used to indicate the probability
 that they are BHB stars. Amongst those with the appropriate $(B-V)_0$ $<$ 0.20, 14  
 have a high probability of being BHB stars
 and  33 are unlikely to be BHB stars. The classification of 8 is ambiguous but
 further study shows that only 4 of these are likely to be BHB stars. 
 There are 10 stars in the 
 $CHSS$ Century survey that are in the $NR$ field; 3 are classified as BHB 
  stars in both our and the $CHSS$ surveys. We classify the other 7 stars
  as unlikely to be BHB stars and they are either given non-BHB types or
 are unclassified in the $CHSS$ survey.  The surface densities (stars 
 deg$^{-2}$) in both our survey of the $NR$ field and that of an adjoining 
 $CHSS$ field show  good agreement in
 completeness between the two surveys. This is not true of the $B2M$ survey 
 where about three quarters of the 26 stars which Beers et al.\ (2007b) classify
  as likely to be BHB stars are classified by us as non-BHB stars.
  Smith et al (2010) have used $SDSS$ colors to assign probabilities for stars
 in the $DR7$ release to be BHB stars. There is fair agreement betwen our
 classifications and these probabilities for the 5 stars in our survey that
 are faint enough to be included in their catalog. A comparison of the
 Smith et al.\ probabilities with a sample from the $CHSS$ survey indicates, 
 however, that the use of the $SDSS$ colors alone can only  isolate a sample 
 that is about 80\% pure.
 
 Among the early-type stars in the $NR$ field,  we identify
 three type $ab$ RR Lyrae variables with mean $V$ magnitudes brighter than 15 
 (the approximate limit of the BHB survey). New photometric and spectroscopic
 data are given for these stars. A fourth type {ab} RR Lyrae has been found
 in the field by Kinemuchi (2006). Its $V$ magnitude is $\sim$15. A fifth 
 type {ab} field RR Lyrae of lower amplitude was found in the field by the $LONEOS$
 survey but was not found by the $NSVS$ because of confusion with brighter
 stars. We estimate from the $LONEOS$ survey that there should be 7 type $ab$
 RR Lyraes in the $NR$ field or about half of the number of BHB stars.

\acknowledgments   
      We thank Dr. Nick Suntzeff for kindly making available the radial
      velocities of 18 early-type stars in the $NR$ survey. 
      This research has made use of the VizieR catalog access tool, CDC,
      Strasbourg, France and 2MASS data provided by the NASA/IPAC Infrared 
      Science Archive, which is operated by the Jet Propulsion Laboratory,
      California Institue of Technology, under contract with NASA.
      Use was also made of MAST (Multimission Archive at the STSci which is
      operated for NASA by AURA), the SIMBAD database (operated at the CDS,
      Strasbourg, France), ADS (the NASA Astrophysics Data System) and the
      Astro-ph e-print server. Finally we would like to thank the referee
      for a careful reading of this text that has enabled us to make a number
      of corrections and clarifications.

\appendix  

\section{The Southern Field $SR$ of Rodgers et al.\ (1993a). }

Fig.\ 14 shows the distribution of galactocentric radial velocities (V$_{gal}$)
 for the early type stars in the $SR$ field (above) and $NR$ field (below). 
 V$_{gal}$ was calculated from the heliocentric radial velocity V$_{hel}$ 
 following Xue et al.\ (2008)\footnote{In this paper we adopt a solar 
 galactocentric distance of 8.0 kpc (Reid \& Brunthaler 2004) but an LSR
 circular velocity velocity of 220 km s$^{-1}$ rather than their somewhat 
 higher value of 236$\pm$15 km s$^{-1}$ to maintain compatibility with other
 recent work.}: 
 \begin{equation}
V_{gal} = V_{hel} + 10\times\cos~l\cos~b + 225\times \sin~l\cos~b +7\times \sin~b    \\ 
 \end{equation}
 The $SR$ field has the same galactic coordinates (l,b) = (270,--45) as the 
 field of Gilmore et al.\ (2002) but does not show  evidence for an
 anomalous disk in its radial velocity distribution. The $SR$ field does,
 however, show a peak around +125 km s$^{-1}$ in its 
 velocity distribution. This shows up as a peak around --30 km s$^{-1}$
 in its galactocentric radial velocity distribution. 
 A similar peak may be present at +30 km s$^{-1}$ in the $NR$ field but it
 is certainly much less pronounced 
  than the peak in the $SR$ field.

\begin{figure}
\centerline{\includegraphics[width = 3.5in]{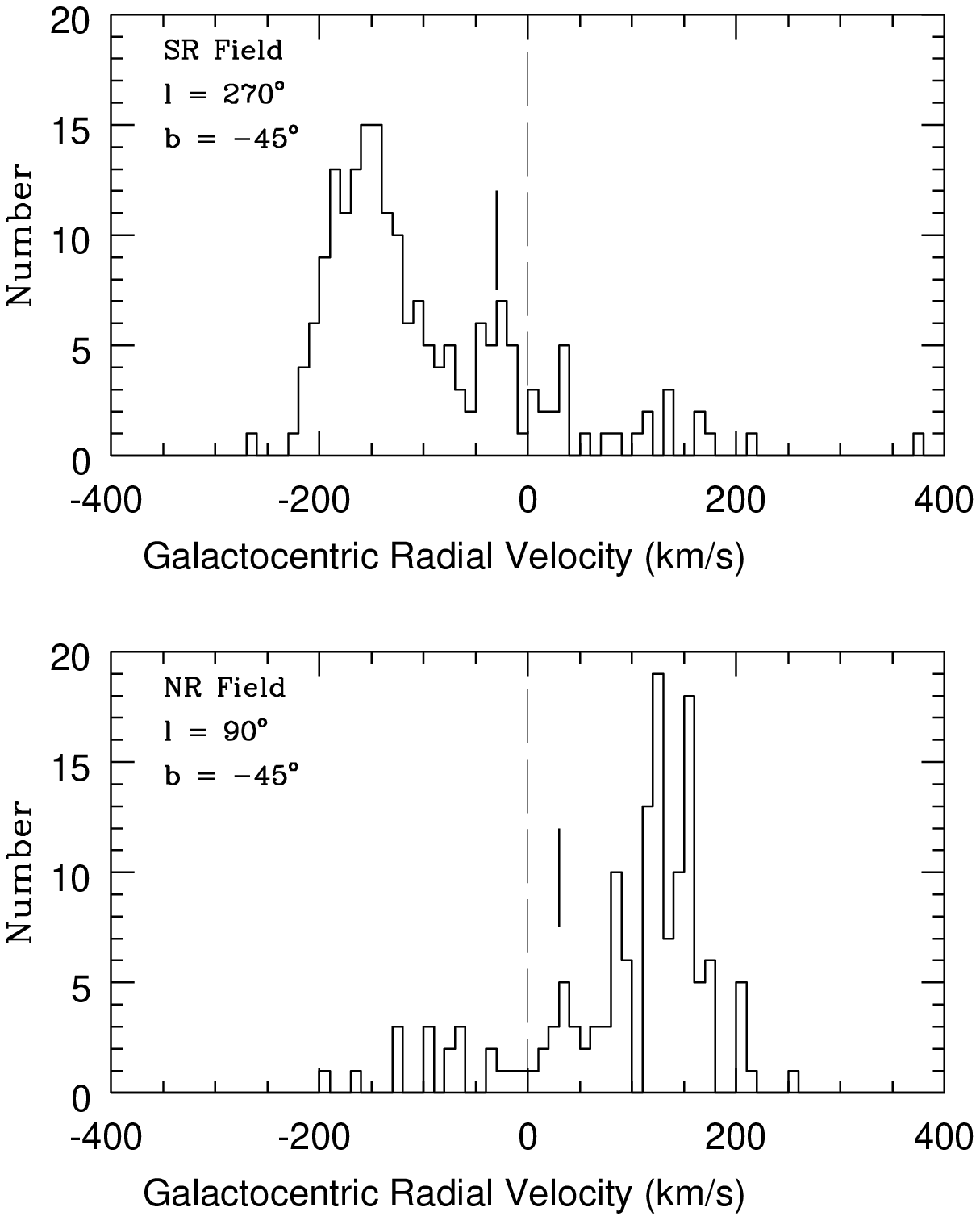}}
\caption{
  Distribution of the galactocentric radial velocities (V$_{gal}$) in km s$^{-1}$
  for the $SR$ (above) and $NR$ (below) fields. The vertical dashed line at
  (V$_{gal}$ = 0.0) is the expected mean velocity for halo stars. The short vertical
  bars show the location of the peak in the $SR$ field and its equivalent
  location in the $NR$ field.
\label{Fig14}}
\end{figure}

 There are 24 stars in the $SR$ field in the velocity range that covers this 
 peak (--05$<$V$_{gal}$$<$--50 km s$^{-1}$). They are listed in Table 8 
 with their proper motions and positions taken from the $UCAC3$
 catalog (Zacharias et al.\ 2010). This table gives the reduced proper motion
 (H) (defined by equation (3) in Sec. 8.2) for each star. We know from our
 study of the BHB stars in the $NR$ field that BHB stars have an H that is
 roughly in the range 8.5 to 11.5. There are 8 to 10 stars in this range 
 (depending on whether one also takes the Ca\,{\sc ii} K- line equivalent width
 into account). This is an upper limit to the number of BHB stars in this
 velocity range because non-BHB stars can also have H in this range. {\it We 
 conclude that the peak at --30 km s$^{-1}$ in V$_{gal}$ must primarily be
 caused by non-BHB stars.}

\begin{deluxetable}{ccccccccccccc}
\tablewidth{0cm}
\tabletypesize{\footnotesize}
\setcounter{table}{8}
\tablecaption{ Stars in Rodgers et al.\ $SR$ Field with Galactocentric Radial Velocities in Range $-$05 to $-$50 km s$^{-1}$.}
\tablehead{ 
\colhead{ No. \tablenotemark{a}  } &
\colhead{  ID \tablenotemark{b}  } &
\colhead{ RA     } &
\colhead{ Dec      } &
\colhead{  RV$_{lsr}$ \tablenotemark{c}  } &
\colhead{  RV$_{gal}$ \tablenotemark{d}  } &
\colhead{ $\mu_{\alpha}$  \tablenotemark{e}  } &
\colhead{ $\mu_{\delta}$ \tablenotemark{f}  } &
\colhead{ H              \tablenotemark{g}  } &
\colhead{  ST \tablenotemark{h}  } &
\colhead{  $V$ \tablenotemark{i}  } &
\colhead{ W$_{0}$(K) \tablenotemark{j}    } &
\colhead{ W(H$\delta$) \tablenotemark{k}    } \\
   &    &(2000J) &(2000J)   & km s$^{-1}$ & km s$^{-1}$ & (mas) & (mas) & & & (mag.) & \AA &   \AA  \\  
}

\startdata
  001& P83l--10 &03:08:48.54 &--67:51:28& +113&  --29& +18.4$\pm$1.8 &--34.0$\pm$ 1.8 &12.3&A2&14.5& 1.4& 11.4 \\  
  011& P83l--16 &03:15:09.98 &--66:07:50& +115&  --27& +04.2$\pm$2.8 &--09.1$\pm$ 2.8 &09.9&A1&15.2& 0.2& 22.4 \\
  027& P83l--18 &03:21:19.12 &--66:18.06& +109&  --35& +07.7$\pm$2.8 & +04.4$\pm$ 2.5 &08.6&A4&14.2& 3.3& 13.6 \\ 
  029& P83s--44 &03:22:34.13 &--63:39:46& +102&  --40& +01.8$\pm$1.7 &--43.4$\pm$ 1.7 &11.4&F3&13.3& 5.6&  6.6 \\ 
  030& P83l--19 &03:22:37.99 &--66:21:33& +123&  --21& +06.5$\pm$2.8 & +01.5$\pm$ 3.4 &08.4&A0&15.0& 1.4& 25.0 \\  
  050& P117l--10&03:26:30.21 &--60:18:49& +120&  --19& +12.0$\pm$13.2&--18.9$\pm$ 2.0&10.5&A0&14.3& 0.0& 17.9  \\
  053& P117l--22&03:26:57.10 &--58:04:26& +107&  --30& +33.4$\pm$2.4 & +23.2$\pm$ 2.5&13.0&A0&15.1& 2.3& 15.3  \\ 
  072& P83l--54 &03:31:38.22 &--64:07:37& +134&  --11& +05.9$\pm$2.5 &--06.0$\pm$ 2.5&09.4&A0&15.1& 0.7& 15.0  \\ 
  077& P117l--12&03:33:28.30 &--59:40:29& +097&  --44& +03.3$\pm$2.3 & +02.1$\pm$ 2.2&06.3&A4&14.6& 3.2& 11.8  \\ 
  080& P117l--30&03:34:34.82 &--57:49:17& +098&  --41& +07.5$\pm$1.8 &--01.1$\pm$ 1.7&09.2&A0&15.0& 0.0& 16.2  \\ 
  081& P83l--77 &03:34:21.13 &--62:05:56& +117&  --27&--00.1$\pm$2.3 &--02.9$\pm$ 1.8&$<$7&A2&15.0& 2.4& 17.5  \\ 
  091& P83l--76 &03:39:22.53 &--63:28:58& +109&  --37& +08.9$\pm$2.0 &--07.4$\pm$ 2.1&10.4&A1&15.2& 1.0& 21.3  \\
  093& P83l--73 &03:42:02.58 &--63:18:57& +111&  --36& +03.8$\pm$2.0 & +04.6$\pm$ 1.6&07.9&A0&14.3& 0.3& 16.9  \\   
  099& P117l--35&03:44:57.82 &--56:41:52& +123&  --19& +06.2$\pm$2.5 &--03.0$\pm$ 6.6&$<$8&A0&14.7& 0.7& 15.4  \\
  106& P117s--20&03:45:56.92 &--61:19:11& +104&  --42& +23.2$\pm$1.2 & +08.8$\pm$ 1.8&08.8&A0&11.9& 0.0& 16.5  \\
  111& P83l--70 &03:47:10.41 &--63:17:27& +125&  --23&--05.4$\pm$1.4 & +00.6$\pm$ 1.4&06.8&A1&13.4& 0.0&  9.4  \\
  129& P117l--44&03:52:18.28 &--59:25:24& +112&  --35& +12.0$\pm$5.8 &--19.7$\pm$ 5.8&10.3&A1&13.7& 0.4& 13.4  \\
  145& P83l--29 &03:58:00.27 &--66:27:09& +104&  --49& +02.2$\pm$4.4 & +00.8$\pm$ 2.5&$<$8&A0&15.3& 0.0& 15.5  \\ 
  147& P83s--23 &03:58:12.55 &--66:29:41& +103&  --50& +20.7$\pm$1.2 & +03.8$\pm$ 1.0&08.4&A4&11.9& 2.1& 16.9 \\
  159& P83l--37 &04:00:43.48 &--64:15:34& +134&  --19& +00.3$\pm$2.2 & +01.1$\pm$ 2.4&$<$7&A0&14.8& 0.6& 16.0 \\
  167& P117l--66&04:04:19.67 &--59:47:24& +130&  --21& +08.0$\pm$1.2 &--00.7$\pm$ 1.5&06.6&A0&12.2& 0.0& 11.9 \\  
  180& P83l--38 &04:05:42.79 &--64:26:47& +142&  --12& +04.8$\pm$1.8 &--02.9$\pm$ 1.8&07.3&A2&13.9& 0.0& 17.2 \\ 
  182& P83l--31 &04:07:05.52 &--65:34:30& +126&  --29& +06.5$\pm$2.5 &--00.6$\pm$ 2.6&08.9&A1&15.3& 0.3& 15.3 \\ 
  184& P83l--36 &04:07:26.79 &--64:44:57& +109&  --46&--01.6$\pm$2.4 & +01.6$\pm$ 7.0&$<$8&A1&15.0& 0.3& 16.3 \\

\enddata
\tablenotetext{a}{Number from Rodgers et al.\ (1993)                        }
\tablenotetext{b}{ID from Rodgers et al.\ (1993)                            }
\tablenotetext{c}{Radial Velocity corrected to LSR from Rodgers et al.\ (1993a).}
\tablenotetext{d}{Galactocentric Radial Velocity.                          }
\tablenotetext{e}{Proper Motion in R.A. from UCAC3 (Zacharias et al.\ 2010)   }
\tablenotetext{f}{Proper Motion in Dec. from UCAC3 (Zacharias et al.\ 2010)   }
\tablenotetext{g}{Reduced Proper Motion (Equation (3) in Sec 8.2).                    }
\tablenotetext{h}{Spectral Type from Rodgers et al.\ (1993a).                }
\tablenotetext{i}{$V$ magnitude from Rodgers et al.\ (1993a).                }
\tablenotetext{j}{Ca\,{\sc ii} K-line Eq. Width corrected for an interstellar
  component according to Rodgers et al.\ (1993a).                }
\tablenotetext{k}{H$\delta$ Eq. Width from Rodgers et al.\ (1993a).                }

\end{deluxetable}

\section{Additional Data for Stars in the $NR$ Field. }
         
\subsection{Improved positions for the fainter stars in the $NR$ Field.}
  The positions given in Rogers et al.\ (1993a) are generally good to a
  few arcsec and therefore adequate for identification. A number of the 
  fainter stars, however, had poorer positions and the coordinates of them
  (taken form the USNO-B 1.0 catalog (Monet et al.\ 2003)) are given in 
   Table 9.

\begin{deluxetable}{cccc}
\tablewidth{0cm}
\tabletypesize{\footnotesize}
\setcounter{table}{9}
\tablecaption{ Coordinates for the Fainter Stars in the $NR$ Field. }
\tablehead{ 
\colhead{  No.  } &
\colhead{  ID   } &
\colhead{ RA (2000J)    } &
\colhead{ Dec (2000J)     } \\
}

\startdata
    18 &    Pn23l1--11 &  23:11:23.8 &   +09:58:46 \\                       
    25 &    Pn23s1--13 &  23:12:21.6 &   +10:47:03 \\                       
    60 &    Pn24l--42  &  23:18:13.1 &   +08:38:08 \\  
    63 &    Pn23l2--2  &  23:18:34.5 &   +13:51:26 \\                       
    72 &    Pn23l2--4  &  23:20:07.8 &   +12:37:47 \\                       
        &      &      &         \\
    75 &    Pn24l--13  &  23:20:47.0 &   +05:11:43 \\                       
    83 &    Pn24l--16  &  23:21:47.7 &   +07:50:56 \\                       
    87 &    Pn23l2--35 &  23:22:20.8 &   +12:06:59 \\                       
    89 &    Pn23l1--28 &  23:22:25.8 &   +11:47:56 \\                       
    92 &    Pn23l1--29 &  23:22:34.1 &   +12:05:08 \\                       
        &      &      &         \\
   104 &    Pn23l2--58 &  23:24:38.2 &   +07:51:02 \\                       
   121 &    Pn23l2--27 &  23:26:25.2 &   +12:46:19 \\                       
   122 &    Pn23l2--55 &  23:26:37.9 &   +08:35:26 \\                       
   123 &    Pn24l--24  &  23:26:37.7 &   +08:36:10 \\                       
   124 &    Pn24l--21  &  23:26:44.6 &   +06:17:42 \\                       
   138 &    Pn24l--5   &  23:28:47.9 &   +05:14:55 \\                       
        &      &      &         \\
\enddata

\end{deluxetable}

\subsection{The $B2M$ Survey.}

 Beers et al.\ (2007b) list a number of BHB candidates in the area of sky of the
 $NR$ survey that are not listed by Rodgers et al.\ (1993a). They are mostly 
 either too faint or too red to have been included in the $NR$ survey; they
 are listed in Table 10.

\begin{deluxetable}{cccccc}
\tablewidth{0cm}
\tabletypesize{\footnotesize}
\setcounter{table}{10}
\tablecaption{BHB candidates in $B2M$ Survey that are not in $NR$ Survey.}
\tablehead{ 
\colhead{  ID \tablenotemark{a}  } &
\colhead{$V_{20}$ \tablenotemark{b}} & 
\colhead{$(B-V)_{20}$ \tablenotemark{c}} & 
\colhead{ RA (2000J)    } &
\colhead{ Dec (2000J)     }& 
\colhead{ Probability\tablenotemark{d}                       } \\
  BPS    & (mag.) & (mag.) &          &         &        \\
}

\startdata
    CS~30338--100 & 13.96 & 0.637 &  23:13:51.5 &   +09:27:25 &   L \\                       
    CS~30338--104 & 15.09 & 0.050 &  23:14:14.5 &   +11:10:33 &   H \\                       
    CS~30338--097 & 13.78 & 0.464 &  23:15:54.0 &   +07:43:14 &   L: \\                       
    CS~30338--098 & 12.13 & 0.490 &  23:17:05.0 &   +07:41:15 &   L \\                       
    CS~30338--091 & 15.54 &--0.020&  23:17:11.4 &   +09:55:49 &   H \\                       
    CS~30338--081 & 13.75 & 0.501 &  23:18:01.9 &   +08:40:17 &   L \\                       
    CS~30338--053 & 14.19 & 0.365 &  23:21:21.4 &   +07:20:35 &   M \\                       
    CS~30338--059 & 13.84 & 0.564 &  23:23:02.7 &   +10:19:28 &   L \\                       
    CS~30338--061 & 14.02 & 0.779 &  23:23:14.9 &   +11:54:30 &   L \\                       
    CS~30338--047 & 12.81 & 0.416 &  23:25:21.8 &   +10:41:15 &   L \\                       
    CS~30338--028 & 14.62 & 0.674 &  23:25:29.3 &   +08:59:50 &   L \\                       
    CS~30338--033 & 14.80 & 0.551 &  23:25:57.6 &   +09:45:54 &   L \\                       
    CS~30338--015 & 15.73 & 0.103 &  23:28:12.3 &   +10:59:47 &   H \\                       
    CS~30338--005 & 12.45 & 0.717 &  23:31:02.7 &   +08:34:51 &   L \\                       
    CS~29522--103 & 13.54 & 0.487 &  23:32:02.1 &   +10:52:13 &   L \\                       
    CS~31088--091 & 14.56 &--0.052&  23:33:58.1 &   +04:03:58 &   H \\                       
    CS~30333--112 & 14.76 & 0.399 &  23:34:34.4 &   +09:08:25 &   M \\                       
    CS~30338--100 & 13.96 & 0.637 &  23:28:47.9 &   +05:14:55 &   L \\                       
\enddata

\tablenotetext{a}{ID from Beers et al.\ (1988)}  
\tablenotetext{b}{Dereddened $V$ magnitude from from Beers et al.\ (2007b)}  
\tablenotetext{c}{Dereddened $(B-V)$ magnitude from from Beers et al.\ (2007b)}  
\tablenotetext{d}{Probability that candidate is BHB star (Beers et al.\ (2007b)  
    (H = High; M = Medium; L = Low)}
\end{deluxetable}

\subsection{Miscellaneous Data for Stars in the $NR$ field from Rodgers 
     et al.\ (1993b).}

 The spectrophotometry and radial velocities of the stars with spectral
types earlier than F0 given in Table 1 of Rodgers et al.\ (1993b) are available
 electronically but may not be easily accessible to all readers. We have 
 therefore reproduced these data for the $NR$ field in Table 11.
 The table also includes metallicities [m/H] for these stars that have been
 recomputed assuming that the interstellar component of the 
  Ca\,{\sc ii} K-line has an 
 equivalent width of 0.3 \AA.

\begin{deluxetable}{cccccccc}
\tablewidth{0cm}
\tabletypesize{\footnotesize}
\setcounter{table}{11}
\tablecaption{ Equivalent Widths, Radial Velocities and Metallicities for Program
 Stars with 0.00$\leq$$(B-V)_{0}$$\leq$0.26. }
\tablehead{ 
\colhead{  No.  } &
\colhead{ Object        } &
\colhead{ W$_{0}$(K) \tablenotemark{a}    } &
\colhead{ W(H$\delta$) \tablenotemark{b}    } &
\colhead{ RV$_{LSR}$   \tablenotemark{c}    } &
\colhead{ $(B-V)_{0}$                    } &
\colhead{ [m/H]    \tablenotemark{d} } &
\colhead{ [m/H]    \tablenotemark{e} }  \\
    &    &\AA  & \AA & km s$^{-1}$ &    &  &  \\
}
\startdata
      002 & Pn23l2--18 &2.8 &18.5& +007  &  0.176 &--1.0 & --0.8  \\
      003 & Pn23s1--15 &0.0 &15.0&--211  &--0.026 &$<$--3.0 & --2.0   \\
      004 & Pn24l--37  &1.7 &17.2& {\bf +020} &  0.131 &--1.4 & --1.2   \\ 
      007 & Pn24l--46  &3.1 &16.5& --074  &  0.245 &--1.3 & --1.1   \\ 
      010 & Pn24l--47  &1.1 &18.3& +020  &  0.140 &--1.9 & --1.6   \\
      015 & Pn24l--32  &2.3 &16.4&--204  &  0.04: & +0.0 &$>$0.0   \\ 
      017 & Pn23s1--2  &2.6 &11.6&  000  &  0.171 &--1.0 & --0.8   \\
      020 & Pn24l--35  &2.6 &15.7& --005  &  0.203 &--1.3 & --1.1   \\ 
      024 & Pn24l--34  &5.6 &15.2& --026  &  0.192 &$>$0.0&$>$0.0    \\ 
      027 & Pn24l--48  &1.6 &18.3& +078  &  0.164 &--1.6 &--1.5     \\ 
      030 & Pn24l--50  &0.00&15,6&--220  &--0.049 &$<$--3.0  &--2.0  \\
      034 & Pn23l2--9  &4.4 &19.6& +003  &  0.237 &--0.5 &--0.3     \\ 
      037 & Pn24l--49  &3.1 &12.9& +021  &  0.245 &--1.3 &--1.1       \\  
      038 & Pn24l--51  &0.0 &16.2&{\bf --164} &  0.039 &$<$--3.0&--2.2     \\ 
      041 & Pn23l2--14 &0.0 &19.8&{\bf --270} &  0.023 &$<$--3.0&--2.2      \\
      048 & Pn24l--45  &0.9 &21.5&{\bf --114}  &  0.020 &--1.2 &--0.8      \\ 
      050 & Pn24l--31  &1.8 &19.1& --089  &  0.155 &--1.5 &--1.3       \\ 
      052 & Pn23l2--63 &3.9 &14.5&--154  &  0.246 &--0.9 &--0.7       \\ 
      053 & Pn24l--30  &1.3 &18.1& --021  &  0.135 &--1.7 &--1.4        \\ 
      059 & Pn24l--43  &5.6 &15.8& --032  &  0.219 &$>$0.0&$>$0.0       \\ 
      060 & Pn24l--42  &0.0 &20.8&{\bf --208} &  0.057 &$<$--3.0&--2.6    \\ 
      063 & Pn23l2--2  &3.0 &20.5&{\bf --038}  &  0.184 &--1.0 &--0.7     \\ 
      065 & Pn24l--52  &3.6 &21.2& +043  &  0.200 &--0.6 &--0.4         \\ 
      072 & Pn23l2--4  &3.6 &19.8&{\bf +009}   &  0.280 &--1.2 &--1.3      \\
      077 & Pn24l--15  &2.3 &20.8& +001  &  0.160 &--1.2 &--0.9         \\ 
      079 & Pn24l--14  &0.0 &13.6& +014  &--0.049 &$<$--3.0 &--2.0  \\ 
      081 & Pn24s--15  &0.5 &17.7& +035  &  0.082 &--2.2 &--1.8         \\ 
      083 & Pn24l--16  &1.2 &25.6&--106  &  0.090 &--1.5 &--1.2         \\ 
      085 & Pn23l2--37 &2.9 &22.9& --027  &  0.102 & +0.0 & +0.0          \\
      087 & Pn23l2--35 &0.0 &19.6&--273  & +0.5:: &$\cdots$ &$\cdots$ \\
      090 & Pn23l2--36 &0.1 &15.5&{\bf --020}  &--0.062 &$<$--3.0 & --1.5  \\
      092 & Pn23l1--29 &1.0 &20.1&--220  &  0.037 &--1.3 &--1.0          \\
      094 & Pn23l2--33 &2.2 & 8.2& --065  &  0.4:  &$\cdots$  &$\cdots$    \\
      095 & Pn23s2--20 &4.6 &18.3& +002  &  0.191 & +0.0 & +0.0          \\ 
      099 & Pn23l2--22 &1.6 &13.6&{\bf --213} &  0.105 &--1.3 &--1.0       \\ 
      100 & Pn23s2--28 &4.0 &15.1& --012  &  0.189 &--0.3 & +0.0         \\
      103 & Pn23s2--17 &2.9 &12.7&  000  &  0.234 &--1.4 &--1.2         \\
      104 & Pn23l2--58 &0.1 &17.2&--265  &  0.114 &$<$--3.0 &--2.7    \\
      106 & Pn24l--10  &3.2 &17.3& +078  &  0.131 &--0.3 & +0.0         \\
      110 & Pn23l2--28 &0.1 &19.7& {\bf --054}&  0.047 &--3.0 &--2.2        \\
      111 & Pn24l--17  &4.9 &13.1& --016  &  0.213 & +0.0 &+0.0         \\
      112 & Pn24s--17  &4.7 &13.1& --019  &  0.256 &--0.4 &--0.2       \\
      113 & Pn24l--2   &3.0 &15.7& --093  &  0.213 &--1.1 &--0.9       \\
      115 & Pn24l--55  &0.3 &17.8&{\bf --132} &  0.042 &--2.4 &--1.8       \\
      116 & Pn24s--3   &2.2 &17.0&  --009  &  0.193 &--1.5 &--1.3        \\
      120 & Pn24l--56  &1.8 &13.4&{\bf --064} &  0.198 &--1.7 &--1.6        \\
      121 & Pn23l2--27 &1.4 &22.4& --019  &  0.124 &--1.6 &--1.4        \\
      127 & Pn24l--19  &3.8 &17.0& --091  &  0.250 &--1.0 &--0.8        \\
      128 & Pn23l2--30 &0.4 &16.5&{\bf +058} &  0.057 &--2.2 &--1.8        \\
      133 & Pn23l2--26 &1.1 &17.8&  --001  &  0.176 &--2.0 &--1.8        \\
      134 & Pn24l--8   &3.0 &15.2&{\bf --126} &  0.148 &--0.6 &--0.4        \\
      136 & Pn24s--8   &0.6 &18.2& +031  &  0.074 &--2.0 &--1.6         \\
      138 & Pn23s1--19 &1.4 &17.3& --009  &  0.101 &--1.4 &--1.2        \\
      140 & Pn23l2--53 &2.8 &16.1& {\bf --042}&  0.265 &--1.5 &--1.3        \\
      141 & Pn24l--22  &1.9 &13.8&{\bf --117} &  0.207 &--1.7 &--1.5        \\
      143 & Pn23l2--39 &1.3 &15.5& +017  &  0.116 &--1.5 &--1.4         \\
      145 & Pn24l--4   &1.4 &16.9&{\bf --150} &  0.027 &--0.5 &--0.2     \\
      146 & Pn24l--27  &2.4 &19.5& {\bf --049}&  0.058 & +0.0 &+0.0       \\
      147 & Pn23l2--25 &1.4 &21.8& --092  &  0.131 &--1.6 &--1.4        \\
      148 & Pn24l--26  &3.7 &20.2& {\bf --046} &  0.089 &$>$0.0&$>$0.0     \\
      151 & Pn24l--6   &1.3 &17.8& {\bf +047} &--0.017 & 0.0  & $>$0.0    \\ 
      156 & Pn23l--40  &1.8 &15.5&  --098 &  0.299 &--2.0 &--1.9        \\ 
      157 & Pn23l1--39 &1.4 &16.2& {\bf --049}&  0.075 &--1.2 &--1.0        \\
      158 & Pn24l--28  &4.1 &19.0& +042  &  0.165 &  0.0 &+0.0         \\
      161 & Pn24s--23  &2.5 &18.1& --036  &  0.234 &--1.5 &--1.3        \\  
      162 & Pn24l--29  &0.0 &17.4& --017  &  0.062 &$<$--3.0&--2.3      \\
      163 & Pn23l2--43 &2.3 &17.0&{\bf +010} &  0.227 &--1.5 &--1.4        \\
      165 & Pn24s--55  &1.3 &13.4&--159  &  0.246 &--2.3 &--2.0        \\
      168 & Pn24s--60  &2.8 &18.5& +028  &  0.158 &--0.8 &--0.6        \\  
      169 & Pn23l2--45 &0.2 &13.9&{\bf --012} &  0.047 &--3.0 &--2.0      \\ 
\enddata

\tablenotetext{a}{ Equivalent width of Ca\,{\sc ii}  K line with 0.6 \AA~ correction 
    for interstellar component from Rogers et al.\ (1993b) Table 1.} 
\tablenotetext{b}{ Equivalent width of H $\delta$ line  
  from Rogers et al.\ (1993b) Table 1.}  
\tablenotetext{c}{ Radial velocity corrected to LSR                    
  from Rogers et al.\ (1993b) Table 1 or our adopted values (boldface).} 
\tablenotetext{d}{Approximate metallicity derived from plot of W(K)$_{0}$ and
  $(B-V)_{0}$ as described in Sec. 6.1; interstellar correction 0.6 \AA.} 
\tablenotetext{e}{Approximate metallicity derived from plot of W(K)$_{0}$ and
  $(B-V)_{0}$ as described in Sec. 6.1; interstellar correction 0.3 \AA.} 
                     
\end{deluxetable}

\section{Comparisons between Surveys for BHB Stars.} 

\begin{figure}
\centerline{\includegraphics[width = 3.0in]{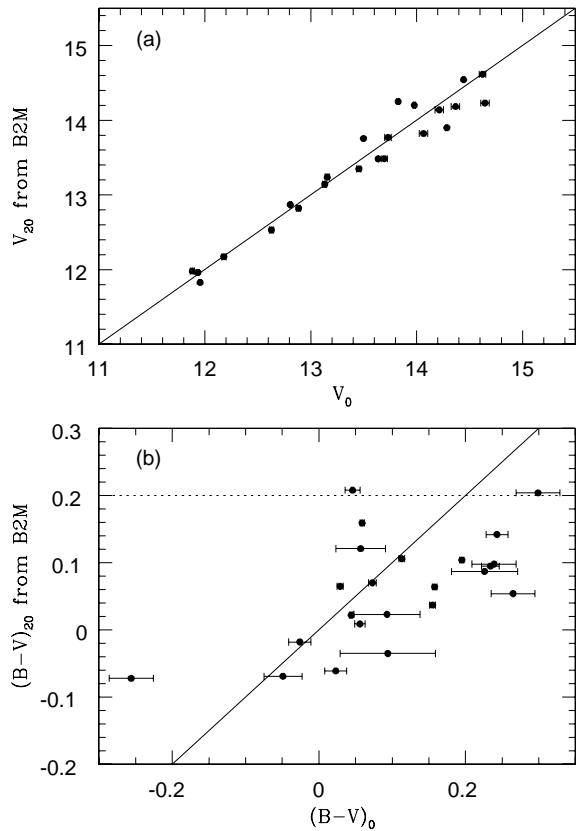}}
\caption{ A comparison of our $V_{0}$ and $(B-V)_{0}$ with the
 $V_{20}$ and $(B-V)_{20}$  given by 
   Beers et al.\ (2007b) in their $B2M$ survey. The horizontal dashed line in 
   Fig. 15 (b) is the $B2M$ red limit of $(B-V)_{20}$ for the star to have a
   high probability of being a BHB star.
\label{Fig 15}}
\end{figure}

\subsection{BHB Stars selected in the $B2M$ Survey.} 
 
 The $B2M$ survey (Beers et al.\ 2007b)  used 2MASS $(JHK)$ magnitudes to derive 
 $V_{20}$ magnitudes and $(B-V)_{20}$ colors for stars taken from 
 the $HK$ objective prism survey of Beers et al.\ (1988, 1996). The stars for
 which they derived  --0.20 $\leq$ $(B-V)_{20}$ $\leq$ +0.20 were given a high  
 (H) probability of being BHB stars.
  Those with +0.20 $\leq$ $(B-V)_{20}$ $\leq$ +0.40 were given a medium (M) 
  probability and redder stars were assigned a low (L) probability.
  The $B2M$ survey overlaps $\sim$50\% of the area of the $NR$ survey.
  We measured $V$ and $(B-V)$ for 23 of the $B2M$ stars in this area. 
   Fig. 15(a) and 15(b) show plots of our $V_{0}$ and $(B-V)_{0}$
  against the corresponding $B2M$ values. The error of 0.14 mag that $B2M$ give
  for their $V_{0}$ magnitudes seems realistic for stars with $V <$~14.0 but may
 be an underestimate for fainter stars. The error of 0.08 mag that $B2M4$ quote
 for their $(B-V)_{20}$ is in good agreement with the error of 0.09 mag that
 we estimate from the scatter in Fig.\ 15(b). Even so, we find that 
 20\% to 25\% of the stars for which  $B2M$ give $(B-V)_{20}$ $\leq$ +0.20 are 
  actually redder than $(B-V)_{0}$ = 0.20 and so are unlikely to be BHB stars.

 The $B2M$ survey has 45 stars in the region of overlap with the $NR$ survey.
    $B2M$ give 26 of these stars a high probability (H) and
 3 a medium probability (M) of being BHB stars. Fig 10(d) shows the 
  $(J-H)_{0}$ $vs.$ $(H-K)_{0}$ plot for the 26 stars that $B2M$ consider 
 having a high (H) 
 probability of being BHB stars. Four of these (41, 56, 122 and 162)
 lie outside the defining window while the color errors of star 87 are too 
 large for it to be selected using 2MASS colors. 
  The remaining 17 stars (those in the sky covered by the $NR$ survey but not
 given in Table 1) are given in Table 10 in Appendix B;
  they are mostly stars that are either too faint or too red to have  
 been discovered by Rodgers et al.\ (1993a).

\subsection{BHB Stars selected in the $CHSS$ Survey.} 

  The Century Halo Star Survey ($CHSS$) (Brown et al. 2008) used $2MASS$ colors to make a
  preliminary selection of BHB candidates and then made a further selection using
  data derived from 
 slit spectra taken with the FAST spectrograph of the Whipple 1.5-m telescope.
 (2.3\AA~resolution, $\lambda\lambda$ 3450---5450 \AA). This additional use of
 spectra not only makes their classification much more secure than that of the
 $B2M$ survey but also provides metallicities and radial velocities. The
 $CHSS$ survey overlaps $\sim$30\% of the $NR$ field and has 12 stars in this
 area. Ten of these are listed in the $NR$ survey and given in Table 1.
  Three of these (CHSS 3068, CHSS 3071 \& CHSS 3075 
  corresponding to stars 99, 110 and 157 in the $NR$) were also classified 
  by them as BHB stars. The other seven were classified otherwise. 
  Of the two $CHSS$ stars in the $NR$ field  
  that were not found by Rodgers et al.\ (1993a), 
 CHSS 3069 is not a BHB star and CHSS 3078 is classed as a BHB star but is too
 faint ($V >$ 15.5) to be in the $NR$ survey.

\subsection{BHB Stars selected by Smith et al.\ (2010) from DR7 of the $SDSS$.}
   DR7 intersects the $NR$ field with Stripes 76 and 79 but only five
 of our BHB candidates are faint enough to be included in this catalog. We 
 classify stars 99 and 110  as BHB3 (likely to be BHB stars) and Smith et al.\
 give these relatively high probabilities of 0.756 and 0.831 of being BHB stars.
 We classify 63, 148 and 163 as  BHB1 (unlikely to be BHB stars);
 these are given lower probabilities of 0.704, 0.261 and 0.123 respectively. 
 Star 63 (for which Smith et al.\ give a probability of 0.704 that it is a BHB
 star)  is CHSS 3064 and is unclassified in the $CHSS$ survey.

\begin{figure}
\centerline{\includegraphics[width = 3.0in]{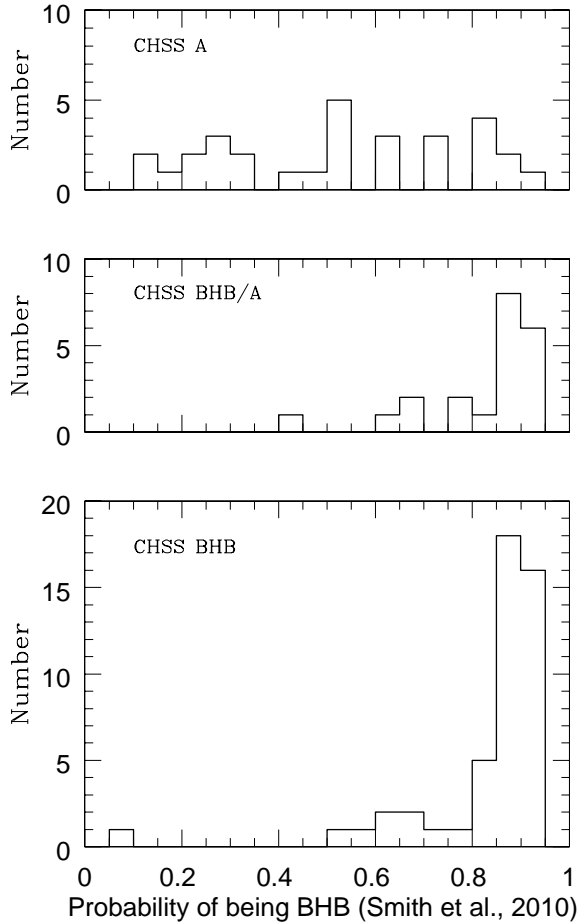}}
\caption{ Histograms of the probabilities of a star being a BHB star 
      according to Smith et al.\ (2010) for various classifications in the
      Century ($CHSS$) Survey: Type A (top); Type BHB/A (middle) and Type 
      BHB (bottom). 
\label{Fig 16}}
\end{figure}

\begin{figure}
\centerline{\includegraphics[width = 3.0in]{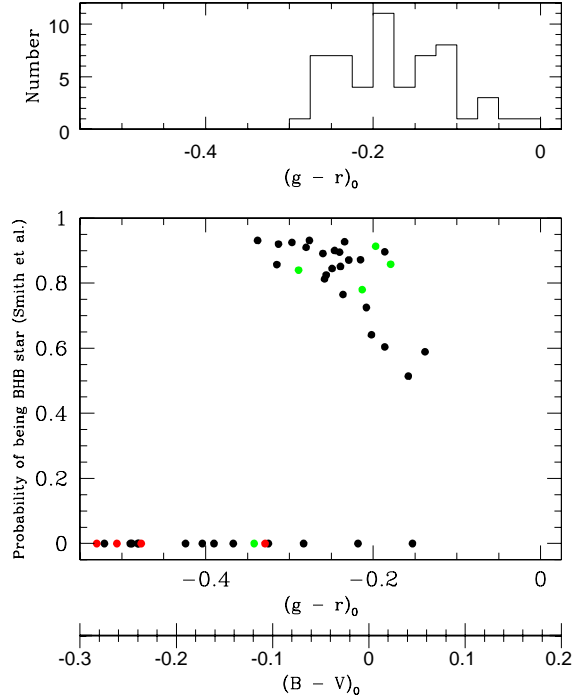}}
\caption{ (above): Color distribution of small random sample of the stars whose
       probability (P) of being a BHB star is greater than 0.8 according to 
      Smith et al.\ (2010). (below): Probability P $vs.$ color $(g-r)_{0}$ for
      BHB stars from the B2M survey with probabilities H (black filled circles);
      H: (red filled circles) and M (green filled circles). B2M stars that are
      not in the Smith et al.\ catalog but are in DR7 have been given P = 0.0.
\label{Fig 17}}
\end{figure}

We compare the $CHSS$ and Smith et al.\ surveys with a sample of stars
with 2MASS $J$ $>$ 15.0. This sample contained 99 stars 
 for which both $CHSS$ classes and Smith 
 et al.\ probabilities (P) were available.  Histograms showing the distribution
of Smith et al.\ probabilities among the $CHSS$ classes are shown in
 Fig 16.\ For the BHB class, 39 stars have P $>$ 0.8 and 9 have P $<$ 0.8. 
  For the BHB/A class, 15 stars have P $>$ 0.8 and 6 have P $<$ 0.8. 
  For the A class, 7 stars have P $>$ 0.8 and 23 have P $<$ 0.8.  By taking the
 ``cut" at P = 0.80, we maximize the agreement for both class BHB and class A;
 in this case we find $\sim$80\% agreement between the two catalog. 
 If we lower 
 the ``cut" to a lower P to increase the agreement with the BHB class we 
 would include more stars that $CHSS$ classify as A class (A-type stars). 

 We make a similar comparison with the B2M catalog for a sample of stars with 2MASS 
 $J$ $>$ 15.0. In Fig.\ 17 the probability (P) from Smith et al.\
 is shown for 
 these stars as a function of the SDSS color $(g-r)_{0}$ with those classed H
 (black), H: (red) and M (green). Of these 46 stars, 21 have P $>$ 0.80 and 18
 have been given  P = 0.0 since they are in the DR7 catalog but not in the
 Smith et al.\ catalog.    Fig.\ 17 also shows the color distribution of a 
 small random  sample of the stars in the Smith et al.\ catalog that have
 P $\geq$ 0.8; they cover the range --0.30 $\leq$ $(g-r)_{0}$$\leq$ 0.00 (equivalent 
 to --0.1 $\leq$$(B-V)_{0}$$\leq$+0.20). The $B2M$ catalog BHB stars have a quite
 different color distribution and are all bluer than $(g-r)_{0}$ = --0.14 
 (or $(B-V)_{0}$ = +0.05). This confirms our earlier conclusion that many of the
 stars in the $B2M$ catalog with an H classification may not be BHB stars.

\section{The Correction for the Interstellar Ca\,{\sc ii} K line.}

\begin{figure}
\centerline{\includegraphics[width = 3.0in]{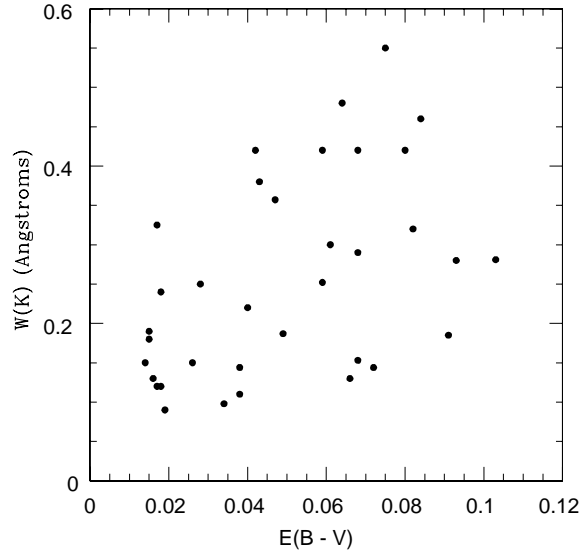}}
\caption{The interstellar Ca\,{\sc ii} K line equivalent width $W(K)$, in \AA,
          {\it vs} the interstellar reddening E$(B-V)$. The sources of the
   data are given in the text.
\label{Fig 18}}
\end{figure}

 The equivalent width  W$_{0}$(K) of 
the Ca\,{\sc ii} K line given by Rodgers et al (1993b) used a constant 
  correction of 0.6 \AA~ for the 
interstellar component. Beers (1990) has given the following 
expression for the strength $W(K)$ of the interstellar K line: 
\begin{equation}  
   W(K) \sin |b| = W_{max} (1 - e^{-|z|/h})   
\end{equation}  
where $W_{max}$ = 0.192 \AA, $z$ is the height in parsecs above the plane, $h$ is a 
scale height (1081 pc) and $b$ is the galactic latitude. According to this expression,
the maximum equivalent width of the interstellar K line at galactic latitude
45$^{\circ}$ is 0.27 \AA. There are six stars in Table 1 of Beers (1990) that
are in the galactic longitude range 75$^{\circ}$ to 105$^{\circ}$ and the 
galactic latitude range --30$^{\circ}$ to --60$^{\circ}$; they have a mean
$W(K)$ of 0.23$\pm$0.06\AA~ and a mean $z$ of 3965 pc. 

We might also expect some correlation between $W(K)$ and the reddening 
E$(B-V)$. Fig.\ 18 is a plot between these quantities for stars whose 
height above the galactic plane is greater than 1 kpc. The W(K) are taken
from Table 1 of Beers (1990) and the E$(B-V)$ from Schlegel et al.\ (1998).
The stars in Table 11 have E$(B-V)$ in the range 0.04 to 0.14 with a mean value
of 0.068.  The correlation between $W(K)$ and E$(B-V)$ is too weak to make it
 worth computing $W(K)$ individually for each E$(B-V)$. On the other hand,
  a mean value of $W(K)$ of 0.3 \AA~is clearly in
 better agreement with the reddening than the 0.6 \AA\ used by Rogers et al.\
 (1993b). We therefore also computed the metallicities 
 for the program stars using a constant $W(K)$ of 0.3 \AA\
 and these are given in the last column of Table 11 in Appendix B.

\clearpage

\begin{deluxetable}{ccccccccccccc}
\tablewidth{0cm}
\tabletypesize{\footnotesize}
\setcounter{table}{1}
\tablecaption{ Photometric Data for Stars in the $NR$ Field of Rodgers et al.\ (1993).}
\tablehead{ 
\colhead{  No.  } &
\colhead{ ID        } &
\colhead{ $V$       } &
\colhead{ $(B-V)$    } &
\colhead{ $(u-B)_{K}$    } &
\colhead{ $\beta$                        } &
\colhead{ E$(B-V)$                            } &
\colhead{ $NUV$      } &
\colhead{ $J$      } &
\colhead{ $K$      } &
\colhead{ $m_{R}$      } &
\colhead{ $\Sigma_{m_{R}}$      } &
\colhead{ Notes    } \\
 (1) & (2) & (3) & (4) & (5) & (6) & (7) & (8)  & (9) & (10)& (11) & (12) & (13)
}

\startdata
    1 &Pn23l2--7  &12.90$\pm$0.02&0.437$\pm$0.010&  ...  &  ...  &       0.09&  16.53&  11.92&  11.74&  13.08&  0.06  &  \\
    2 &Pn23l2--18 &12.27$\pm$0.03&0.238$\pm$0.009&1.934$\pm$0.011&2.843$\pm$0.016&       0.04&  15.09&  11.74&  11.67&  12.45&  0.03  &  \\
    3 &Pn23s1--15 &13.28$\pm$0.02&0.011$\pm$0.015&1.660$\pm$0.019&  ...  &       0.04&   ... &  13.25&  13.28&   ... &  ...   &  \\
    4 &Pn24l--37  &13.22$\pm$0.02&0.194$\pm$0.006&2.183$\pm$0.060&2.820$\pm$0.011&       0.05&  16.08&  12.81&  12.75&  13.55&  0.07  &  \\
    5 &Pn23s1--10 &11.82$^{a}$&   0.35:&  ...  &  ...  &       0.07&  15.20&  11.10&  10.91&  12.16&  0.03  &  \\
    6 &Pn24s--32  &10.63$^{a}$&   0.41:&  ...  &  ...  &       0.04&   ... &   9.81&   9.58&  10.98&  0.01  &  \\
    7 &Pn24l--46  &14.55$\pm$0.01&0.348$\pm$0.025&  ...  &  ...  &       0.09&  17.74&  13.75&  13.51&  14.74&  0.14  &  \\
    8 &Pn23l2--17 &13.23$\pm$0.01&0.402$\pm$0.017&1.930$\pm$0.011&  ...  &       0.05&  16.63&  12.44&  12.25&  13.12&  0.04  &   \\
    9 &Pn24l--38  &13.44$^{a}$&   0.32:&  ...  &  ...  &      0.05&  17.08&  12.77&  12.61&  13.53&  0.07  &   \\
   10 &Pn24l--47  &13.07$\pm$0.01&0.195$\pm$0.012&1.910$\pm$0.026&  ...  &      0.05&  15.62&  12.70&  12.62&  13.22&  0.06  &   \\
   11 &Pn23s2--2  &12.15$^{a}$&   0.46:&  ...  &  ...  &       0.08&  15.88&  11.21&  10.98&  12.41&  0.04  &  \\
   12 &Pn23l2--19 &13.58$\pm$0.01&0.334$\pm$0.045&1.730$\pm$0.039&  ...  &       0.05&  16.68&  13.18&  12.97&  14.02&  0.17  & {\bf 1}  \\
   13 &Pn24l--36  &  13.44&   0.434&  ...  &  ...  &       0.06&  16.29&  12.38&  12.10&  13.61&  0.06  &   \\
   14 &Pn23l2--16 &13.89$^{a}$&0.51:   &  ...  &  ...  &       0.04&  17.39&  12.92&  12.62&  13.94&  0.07  &  \\
   15 &Pn24l--32  &15.12$\pm$0.03&0.118$\pm$0.068&2.255$\pm$0.008&  ...  &       0.08&  17.70&  14.76&  14.75&   9.00&  9.00  &   \\
   16 &Pn23l2--1  &14.02$\pm$0.02&0.476$\pm$0.032&1.870$\pm$0.106&  ...  &       0.07&  17.58&  12.76&  12.52&  14.02&  0.31  & {\bf 2}  \\
   17 &Pn23s1--2  &12.52$\pm$0.01&0.241$\pm$0.006&1.965$\pm$0.014&  ...  &      0.06&  15.51&  12.01&  11.91&  12.80&  0.04  &   \\
   18 &Pn23l1--11 &15.57& --0.230&  ...  &  ...  &     0.05&  ...  &15.54 &15.22  &  15.33&  0.20  & {\bf 3} \\
   19 &Pn23l2--6  &13.84$\pm$0.03&0.447$\pm$0.021&  ...  &  ...  &       0.07&  17.44&  12.82&  12.57&  14.02&  0.08  &  \\
   20 &Pn24l--35  &13.91$\pm$0.02&0.257$\pm$0.016&1.972$\pm$0.030&  ...  &       0.07&   ... &  13.31&  13.17&  14.14&  0.11&    \\
   21 &Pn23s1--3  &12.66$^{a}$&   0.46: &  ...  &  ...  &       0.07&  16.64&  11.67&  11.44&  12.86&  0.03&    \\
   22 &Pn23l2--8  &14.24$^{a}$&   0.33: &  ...  &  ...  &       0.06&  17.69&  13.55&  13.37&  14.64&  0.12&    \\
   23 &Pn23s1--14 &13.22$^{a}$&   0.52: &  ...  &  ...  &       0.04&   ... &  12.18&  11.90&  13.14&  0.05&    \\
   24 &Pn24l--34  &14.31$\pm$0.01&0.269$\pm$0.034&1.811$\pm$0.040&  ...  &       0.08&  17.35&  13.72&  13.56&  14.48&  0.11&    \\
   25 &Pn23s1--13 &13.10$^{a}$& --0.260&  ...  &  ...  &      0.04&  ...  &  13.82&  14.08&  13.77&  0.08& {\bf 4}   \\
   26 &Pn24l--33  &14.38$\pm$0.04&0.353$\pm$0.065&  ...  &  ...  &       0.08&  17.57&  13.42&  13.16&  14.52&  0.14&   \\
   27 &Pn24l--48  &12.77$\pm$0.01&0.212$\pm$0.012&1.938$\pm$0.070&  ...  &       0.05&  15.55&  12.31&  12.17&  13.12&  0.04&    \\
   28 &Pn24s--33  &12.42$\pm$0.01&0.420$\pm$0.003&1.978$\pm$0.004&  ...  &       0.11&  16.34&  11.51&  11.31&  12.61&  0.03&   \\
   29 &Pn23l2--5  &13.02$^{a}$&   0.56: &  ...  &  ...  &       0.05&  15.91&  11.91&  11.60&  13.22&  0.04&   \\
   30 &Pn24l--50  &14.70$\pm$0.02&0.007$\pm$0.022&1.821$\pm$0.116&  ...  &       0.06&  16.61&  14.54&  14.48&  14.80&  0.15&      \\
   31 &Pn24s--31  &12.69$\pm$0.02&0.423$\pm$0.011&  1.752$\pm$0.042&  ...  &       0.06&  16.47&  11.81&  11.58&  12.76&  0.06&   \\
   32 &Pn23l2--20 &  14.10$^{a}$&   0.66: &  ...  &  ...  &       0.04&  ...  &  12.98&  12.55&  13.97&  0.09&   \\
   33 &Pn23s1--4  &11.11$\pm$0.04&0.348$\pm$0.003&1.912$\pm$0.020&2.755$\pm$0.010&       0.05&  14.70&  10.40&  10.24&  11.43&  0.02&   \\
   34 &Pn23l2--9  &13.38$\pm$0.02&0.313$\pm$0.015&2.040$\pm$0.014&  ...  &       0.08&  16.88&  12.74&  12.58&  13.58&  0.06&   \\
   35 &Pn24s--30  &11.95$\pm$0.01&0.408$\pm$0.002&  ...  &  ...  &       0.07&  ...  &  11.13&  10.93&  12.12&  0.02&   \\
   36 &Pn24s--37  &11.92$\pm$0.01&0.398$\pm$0.006&  ...  &  ...  &       0.08&  ...  &  11.15&  10.95&  11.50&  0.04&   \\
   37 &Pn24l--49  &11.54$\pm$0.01&0.298$\pm$0.006&  ...  &  ...  &       0.05&  14.65&  11.17&  11.03&  11.96&  0.14& {\bf 5}  \\
   38 &Pn24l--51  &12.93$\pm$0.01&0.103$\pm$0.004&2.147$\pm$0.036&2.850$\pm$0.024&       0.06&  15.44&  12.60&  12.53&  12.90&  0.04&       \\
   39 &Pn24l--41  &  12.33$^{a}$&   0.43: &  ...  &  ...  &       0.10&  15.85&  11.41&  11.20&  12.55&  0.04&   \\
   40 &Pn23s2--9  &11.76$\pm$0.01&0.330$\pm$0.006&1.973$\pm$0.003&  ...  &       0.05&  15.35&  11.09&  10.98&  12.09&  0.02&   \\
   41 &Pn23l2--14 &14.44$\pm$0.01&0.072$\pm$0.015&2.070$\pm$0.059&  ...  &       0.05&  16.68&  14.15&  14.09&  14.73&  0.13&      \\
   42 &Pn24s--35  &12.18$\pm$0.01&0.418$\pm$0.006&  ...  &  ...  &       0.09&  15.97&  11.35&  11.15&  12.48&  0.03&   \\
   43 &Pn24s--29  &  11.23$^{a}$&   0.28: &  ...  &2.816$\pm$0.010  &       0.07&  ...  &  10.50&  10.40&  11.60&  0.02&   \\
   44 &Pn23l2--69 &  13.52$^{a}$&   0.41: &  ...  &  ...  &       0.06&  16.78&  12.65&  12.43&  13.52&  0.05&   \\
   45 &Pn24s--40  &  11.28$^{a}$&   0.36: &  ...  &2.761$\pm$0.010  &       0.08&  14.63&  10.53&  10.34&  11.43&  0.04&   \\
   46 &Pn23s2--10 &  10.59$^{a}$&   0.32: &  ...  &2.742$\pm$0.005  &       0.05&  14.12&   9.90&   9.74&  10.96&  0.02&   \\
   47 &Pn23s2--48 &  12.42$^{a}$&   0.52: &  ...  &  ...  &       0.05&  15.98&  11.32&  11.06&  12.68&  0.12& {\bf 6}  \\
   48 &Pn24l--45  &14.63$\pm$0.01&0.085$\pm$0.004&2.154$\pm$0.078&  ...  &       0.06&  ...  &  14.37&  14.40&  ...  &  ... &    \\
   49 &Pn23s2--11 &10.41$\pm$0.01&0.316$\pm$0.002&1.858$\pm$0.007&2.768$\pm$0.008&       0.04&  ...  &   9.74&   9.59&  10.77&  0.02&    \\
   50 &Pn24l--31  &13.05$\pm$0.03&0.225$\pm$0.021&1.878$\pm$0.030&  ...  &       0.07&  15.49&  12.54&  12.37&  13.35&  0.07&   \\
   51 &Pn23l2--13 &  13.14$^{a}$&   0.54: &  ...  &  ...  &       0.05&  16.23&  12.04&  11.75&  13.27&  0.06&   \\
   52 &Pn23l2--63 &14.19$\pm$0.04&0.288$\pm$0.030&1.705$\pm$0.040&  ...  &      0.04&  16.89&  13.53&  13.35&  14.60&  0.13&   \\
   53 &Pn24l--30  &13.82$^{a}$&0.189$\pm$0.009&2.013$\pm$0.050&  ...  &       0.07&  16.70&  13.37&  13.29&  14.14&  0.09&   \\
   54 &Pn23l2--10 &  13.36$^{a}$&   0.53: &  ...  &  ...  &       0.05&  16.78&  12.41&  12.08&  13.58&  0.07&   \\
   55 &Pn23l2--11 &  14.95$^{b}$&   0.6:: &  ...  &  ...  &       0.04&  18.58&  13.94&  13.56&  15.36&  0.16&   \\
   56 &Pn24l--44  &  13.75& --0.175&  ...  &  ...  &       0.09&   ... &  14.14&  14.36&  13.34&  0.07& {\bf 7}  \\
   57 &Pn23s2--13 &  12.70$^{a}$&   0.44: &  ...  &  ...  &       0.04&  16.27&  11.78&  11.56&  12.87&  0.03&   \\
   58 &Pn24s--26  &  14.16$^{a}$&   0.45: &  ...  &  ...  &       0.14&   ... &  13.33&  13.05&  14.33&  0.11&   \\
   59 &Pn24l--43  &14.55$\pm$0.01&0.288$\pm$0.009&1.955$\pm$0.040&  ...  &       0.07&  17.89&  13.95&  13.71&  15.03&  0.17&   \\
   60 &Pn24l--42  &14.84$\pm$0.03&0.127$\pm$0.34&2.240$\pm$0.051&  ...  &       0.07&  17.67&  14.32&  14.09&  15.31&  0.20&   \\
\enddata
\tablenotetext{a}{Magnitude from $ASAS-3$ catalog. }
\tablenotetext{b}{Magnitude from $ASAS-3$ and $GSC-2.3$ catalogs.} 
\tablenotetext{(1)}{Variable. See Table 2.}
\tablenotetext{(2)}{RR Lyrae variable IX PEG. See Table 2.}
\tablenotetext{(3)}{Subdwarf KUV~23089+0942. $BV$ from Wegner et al.\ (1990).}
\tablenotetext{(4)}{White Dwarf PG2309+105. $BV$ from Eggen (1968).}
\tablenotetext{(5)}{Variable. See Table 2.}
\tablenotetext{(6)}{Variable. See Table 2.}
\tablenotetext{(7)}{Subdwarf PG2314+076. $BV$ from Allard et al.\ (1994).}
\tablecomments{Only a portion of this table is shown here to demonstrate its form and content.  The full table is available in the published paper.}
\end{deluxetable}

\end{document}